\documentstyle[12pt,axodraw]{article}
\begin{document}
\baselineskip 0.6cm

\def\simgt{\mathrel{\lower2.5pt\vbox{\lineskip=0pt\baselineskip=0pt
           \hbox{$>$}\hbox{$\sim$}}}}
\def\simlt{\mathrel{\lower2.5pt\vbox{\lineskip=0pt\baselineskip=0pt
           \hbox{$<$}\hbox{$\sim$}}}}

\begin{titlepage}

\begin{flushright}
UCB-PTH-05/11 \\
LBNL-57427
\end{flushright}

\vskip 1.8cm

\begin{center}

{\Large \bf 
A Minimally Fine-Tuned Supersymmetric \\ Standard Model
}

\vskip 1.0cm

{\large
Z. Chacko$^a$, Yasunori Nomura$^{b,c}$ and David Tucker-Smith$^d$
}

\vskip 0.4cm

$^a$ {\it Department of Physics, University of Arizona,
                Tucson, AZ 85721} \\
$^b$ {\it Department of Physics, University of California,
                Berkeley, CA 94720} \\
$^c$ {\it Theoretical Physics Group, Lawrence Berkeley National Laboratory,
                Berkeley, CA 94720} \\
$^d$ {\it Department of Physics, Williams College, 
                Williamstown, MA 01267}

\vskip 1.2cm

\abstract{
We construct supersymmetric theories in which the correct scale for 
electroweak symmetry breaking is obtained without significant fine-tuning. 
We calculate the fine-tuning parameter for these theories to be at 
the $20\%$ level, which is significantly better than in conventional 
supersymmetry breaking scenarios.  Supersymmetry breaking occurs 
at a low scale of order $100~{\rm TeV}$, and is transmitted to the 
supersymmetric standard-model sector through standard-model gauge 
interactions.  The Higgs sector contains two Higgs doublets and a singlet 
field, with a superpotential that takes the most general form allowed 
by gauge invariance.  An explicit model is constructed in 5D warped space 
with supersymmetry broken on the infrared brane.  We perform a detailed 
analysis of electroweak symmetry breaking for this model, and demonstrate 
that the fine-tuning is in fact reduced.  A new candidate for dark 
matter is also proposed, which arises from the extended Higgs sector 
of the model.  Finally, we discuss a purely 4D theory which may also 
significantly reduce fine-tuning. 
}

\end{center}
\end{titlepage}

\section{Introduction}
\label{sec:intro}

Low-energy supersymmetry is an attractive framework for physics beyond 
the standard model.  It not only stabilizes the electroweak scale 
against potentially large radiative corrections, but also leads to a 
successful prediction for the low-energy gauge couplings through gauge 
coupling unification~\cite{Dimopoulos:1981zb}.  However, non-discovery 
of both superparticles and a light Higgs boson at LEP~II puts strong 
constraints on theories with low-energy supersymmetry.  To avoid the 
constraints from LEP~II, especially those on the physical Higgs-boson 
mass, masses of superparticles must generically be pushed up to larger 
values.  This then leads to a large negative Higgs-boson mass-squared 
parameter at radiative level, requiring fine-tuning among parameters 
to reproduce the correct scale for electroweak symmetry breaking. 

In this paper we construct a supersymmetric theory that does not suffer 
from significant fine-tuning.  We do this without spoiling attractive 
features of supersymmetry such as the successful prediction associated with 
gauge coupling unification.  We consider the case in which all the states 
in the minimal supersymmetric standard model (MSSM) are elementary up to 
high energies of order the unification scale $\simeq 10^{16}~{\rm GeV}$. 
We find that, to evade severe fine-tuning, an additional contribution 
to the physical Higgs-boson mass is needed, beyond those in the MSSM. 
We provide it by coupling the two Higgs doublets to a singlet superfield 
$S$, as in the case of the next-to-minimal supersymmetric standard model 
(NMSSM)~\cite{Nilles:1982dy}.  The form of the superpotential in the 
Higgs sector of our model, however, is not identical to that of the 
NMSSM, allowing parameter regions that are not available in the NMSSM 
Higgs potential.  The mass of the physical Higgs boson can also be heavier 
than in the simplest NMSSM due to additional contributions to the evolution 
of couplings arising from the sector that dynamically breaks supersymmetry. 
Raising the Higgs-boson mass, however, is not sufficient to give a 
significant reduction of fine-tuning.  We find that reducing the 
fine-tuning also requires a low mediation scale for supersymmetry breaking, 
and a superparticle spectrum that does not respect simple unified mass 
relations.  Our theory naturally accommodates all these features.

The superpotential of the Higgs sector in our theory is effectively 
given by
\begin{equation}
  W_{\rm Higgs} = \lambda S H_u H_d + L_S^2 S 
    + \frac{M_S}{2}S^2 + \frac{\kappa}{3}S^3,
\label{eq:intro-Higgs}
\end{equation}
where $H_u$ and $H_d$ are the two Higgs doublets of the MSSM, 
$\lambda$ and $\kappa$ are dimensionless coupling constants, and $L_S$ 
and $M_S$ are dimensionful parameters of order the electroweak scale. 
This superpotential can be obtained by integrating out a set of 
singlet fields, collectively called $P$ and $X$, that couple both 
with the $S$ field and with the dynamical supersymmetry breaking (DSB) 
sector.  Due to interactions with the DSB sector, these singlet 
fields feel the scale of dynamical supersymmetry breaking, which is 
of order $100~{\rm TeV}$ in our theory, and generate mass parameters 
of order the weak scale, such as $L_S$ and $M_S$, in the Higgs sector. 
We find that some of the singlet fields ($X$ fields) naturally develop 
vacuum expectation values (VEVs) of order the weak scale, while 
others ($P$ fields) do not.  Because the masses of the singlet fields 
are naturally of order the weak scale, they may affect phenomenology 
at the electroweak scale.  Moreover, our theory possesses an unbroken 
$Z_2$ symmetry, under which the $P$ fields, which we call pedestrian 
fields, are odd and all other fields are even.  This makes the 
lightest particle in the $P$ multiplets a good candidate for 
the dark matter of the universe. 

To explicitly realize a theory incorporating all the features 
described above, we employ a warped space construction, in which 
dynamical supersymmetry breaking is described as supersymmetry 
breaking on the infrared brane in 5D warped space truncated 
by two branes.  In particular, our model employs the basic 
structure of the model constructed in Ref.~\cite{Nomura:2004is} --- 
the gauge group in the bulk of the truncated 5D AdS space is $SU(5)$ 
but it is broken to the $SU(3)_C \times SU(2)_L \times U(1)_Y$ 
(321) subgroup both at the ultraviolet and infrared branes.  One 
of the virtues of constructing models in warped space is that 
they provide calculable theories of dynamical supersymmetry 
breaking~\cite{Nomura:2004zs}.  In the present context, we can 
explicitly quantify the degree of fine-tuning of our theory by 
studying the fractional change of the weak scale in response to 
fractional changes of fundamental parameters $a_i$ of the theory, 
$\Delta^{-1} = \min_i|(a_i/M_Z^2)(\partial M_Z^2/\partial 
a_i)|$~\cite{Barbieri:1987fn}.  We find that the required 
``fine-tuning'' in our theory is quite mild,
\begin{equation}
  \Delta^{-1} \simeq 20\%.
\end{equation}
This is a significant improvement over conventional supersymmetry 
breaking scenarios such as supergravity mediation, which typically 
require a fine-tuning of $\Delta^{-1} \simeq (2\!\sim\!3)\%$ or 
even worse. 

The organization of the paper is as follows.  In the next section 
we discuss sources of fine-tuning in supersymmetric theories, and 
motivate the particular construction we adopt in later sections. 
In section~\ref{sec:tuningless} we discuss the basic structure 
of our theory, and provide details of its Higgs sector.  In 
section~\ref{sec:theory} we present an explicit model constructed 
in 5D warped space, including a discussion of the singlet sector. 
Electroweak symmetry breaking is studied in section~\ref{sec:ewsb}, 
where we perform a renormalization group analysis, minimize the Higgs 
potential, and find that the fine-tuning can be significantly reduced 
in our model.  We also calculate the spectrum of superparticles for 
a few sample points in the parameter space.  Some phenomenological 
issues, including pedestrian dark matter, are discussed in 
section~\ref{sec:pheno}.  In section~\ref{sec:alternative} we discuss 
an alternative class of theories, constructed purely in 4D, which 
also may significantly reduce fine-tuning.  Finally, our conclusions 
are given in section~\ref{sec:concl}.

For recent, alternative ideas to address the supersymmetric 
fine-tuning problem, see for example Refs.~[\ref{Bastero-Gil:2000bw:X}%
~--~\ref{Babu:2004xg:X}].

\section{Sources of the Supersymmetric Fine-Tuning Problem}
\label{sec:sources}

In this section we discuss possible sources of fine-tuning in generic 
classes of supersymmetric theories.  The leading contribution to the 
negative Higgs-boson mass-squared parameter in the supersymmetric 
standard model (SSM) comes from loops of top quark and squarks.  This 
contribution is given approximately by
\begin{equation}
  m_h^2 \simeq -\frac{3y_t^2}{4\pi^2}\, m_{\tilde{t}}^2\, 
    \ln\Biggl( \frac{M_{\rm mess}}{m_{\tilde{t}}} \Biggr),
\label{eq:corr-Higgs}
\end{equation}
where $y_t$ is the top Yukawa coupling, $m_{\tilde{t}}$ represents the 
masses of two top squarks, which we have taken to be equal for simplicity, 
and $M_{\rm mess}$ is the scale at which superparticle masses are generated 
(or at which supersymmetry breaking is mediated to the SSM sector).  If 
supersymmetry breaking is mediated at a high scale, this gives a large 
contribution to $m_h^2$.  For example, in the minimal supergravity scenario, 
$M_{\rm mess} \simeq M_{\rm Pl}$, so that Eq.~(\ref{eq:corr-Higgs}) gives 
$-m_h^2$ as large as $(500~{\rm GeV})^2$ even for $m_{\tilde{t}}^2 \simeq 
(300~{\rm GeV})^2$.  In fact, the value of $m_{\tilde{t}}$ should typically 
be larger to obtain sufficiently large Higgs-boson and superparticle masses, 
as discussed later.  This leads to fine-tuning for electroweak symmetry 
breaking, as the electroweak scale is determined at tree level by the equation
\begin{equation}
  \frac{M_Z^2}{2} \simeq -m_h^2 - |\mu|^2,
\label{eq:ewsb-cond}
\end{equation}
where $\mu$ is the supersymmetric mass for the two Higgs doublets.  The 
cancellation required between the two independent parameters $m_h^2$ and $\mu$ 
is, therefore, at the level of a few percent or worse.  To perform a precise 
analysis, however, we must use renormalization group (RG) equations because 
the logarithm appearing in Eq.~(\ref{eq:corr-Higgs}) is large.  Such an 
analysis can be found, for example, in Refs.~\cite{Chankowski:1998xv} for 
the case of the minimal supergravity scenario~\cite{Chamseddine:1982jx}, 
giving fine-tuning at the $(2\!\sim\!3)\%$ level or worse.  Note that 
Eqs.~(\ref{eq:corr-Higgs}) and (\ref{eq:ewsb-cond}) are valid for 
moderately large values for $\tan\beta \equiv \langle h_u \rangle/\langle 
h_d \rangle$, e.g. $2 \simlt \tan\beta \simlt 40$, where $h_u$ and $h_d$ 
are the two Higgs doublets giving masses for the up-type and down-type 
quarks, respectively. 

While it is possible that the fine-tuning described above may just be an 
artifact arising from some special relation among parameters derived in 
some fundamental theory, the simplest possibility for reducing fine-tuning 
is to make the logarithm in Eq.~(\ref{eq:corr-Higgs}) smaller, i.e. to 
lower the mediation scale of supersymmetry breaking.  The extent to which 
this helps depends on the top-squark mass $m_{\tilde{t}}$, and the mediation 
scale $M_{\rm mess}$.  As we are about to see, just making the logarithm 
small is not enough to eliminate fine-tuning in the simplest supersymmetric 
theories. 

In the minimal supersymmetric standard model (MSSM), the non-discovery of 
the Higgs boson at LEP~II requires the mass of the lightest $CP$-even Higgs 
boson, $M_{\rm Higgs}$, to be larger than $114~{\rm GeV}$ in most of the 
parameter space~\cite{unknown:2001xx}.  Since the value of $M_{\rm Higgs}$ 
at tree level is bounded from above by $M_Z$, this requires a significant 
radiative contribution to $M_{\rm Higgs}$.  Such a contribution arises 
from top quark and squark loops, whose dominant piece is proportional 
to $\ln(m_{\tilde{t}}/m_t)$~\cite{Okada:1990vk}. The experimental bound 
$M_{\rm Higgs} \simgt 114~{\rm GeV}$ then implies that the top-squark masses, 
$m_{\tilde{t}}$, should be larger than about $(800\!\sim\!1000)~{\rm GeV}$ 
for relatively small top-squark mixing and about $(500\!\sim\!600)~{\rm GeV}$ 
even if $M_{\rm Higgs}$ is maximized by allowing a large top-squark 
mixing~\cite{Carena:1995wu}.  We thus find that the constraint from the 
Higgs-boson search alone is enough to make the fine-tuning (much) worse 
than $10\%$, especially for the case of a small top-squark mixing, even 
if the logarithm in Eq.~(\ref{eq:corr-Higgs}) is just a factor of two 
or so (corresponding to $M_{\rm mess}$ of order TeV). 

Another constraint on $m_{\tilde{t}}$ comes from the non-discovery of 
superparticles at LEP~II.  This constraint, obviously, depends on the model 
we consider, because it depends on the spectrum of the superparticles. 
What types of models should we consider?  Since generic spectra for the 
superparticles lead to the supersymmetric flavor problem, the mediation of 
supersymmetry must be flavor universal.  For small values of $M_{\rm mess}$, 
and preserving the supersymmetric desert, the most natural mechanism that 
gives flavor-universal superparticle masses is to mediate supersymmetry 
breaking through standard-model gauge interactions (this includes, for 
example, gauge mediation models~\cite{Dine:1981gu,Dine:1994vc} and 
models in warped space with supersymmetry breaking mediated by gauge 
interactions~[\ref{Gherghetta:2000qt:X}~--~\ref{Gherghetta:2000kr:X},~%
\ref{Nomura:2004is:X}]).  Imagine, then, that supersymmetry is dynamically 
broken in a sector respecting a global $SU(5)$ symmetry ($\supset SU(3)_C 
\times SU(2)_L \times U(1)_Y$) and that the breaking is mediated to the 
SSM sector by standard-model gauge interactions.  This is a natural 
assumption because the supersymmetry breaking sector should not disturb 
the success of the gauge coupling unification prediction, and the 
simplest possibility is that this sector respects $SU(5)$.  In this 
case the ratio of the top-squark mass to the right-handed selectron 
mass will be about
\begin{equation}
  \frac{m_{\tilde{t}}^2}{m_{\tilde{e}}^2} 
    \simeq \frac{(4/3)g_3^4}{(3/5)g_1^4} 
    \simeq (7\!\sim\!9)^2,
\label{eq:te-ratio}
\end{equation}
where $g_3$ is the $SU(3)_C$ gauge coupling and $g_1$ is the $U(1)_Y$ gauge 
coupling in the $SU(5)$ normalization, both renormalized at the scale of 
order $M_{\rm mess}$.%
\footnote{This ratio could even be larger because the running contributions 
to $m_{\tilde{t}}^2$ and $m_{\tilde{e}}^2$ below $M_{\rm mess}$, $\delta 
m_{\tilde{t}}^2$ and $\delta m_{\tilde{e}}^2$, typically have a larger 
ratio $\delta m_{\tilde{t}}^2/\delta m_{\tilde{e}}^2 \simeq (4/3)g_3^2 
M_3^2/(3/5)g_1^2 M_1^2 \simeq (4/3)g_3^6/(3/5)g_1^6 \simeq (16\!\sim\!24)^2$, 
where $M_3$ and $M_1$ are the gluino and bino masses, respectively, and 
the quantities appearing in the equation are evaluated at the scale 
$M_{\rm mess}$.  This effect, however, is not very large for small 
values of $M_{\rm mess} = O(1\!\sim\!100~{\rm TeV})$.}
The non-discovery of the right-handed selectron at LEP~II pushes up its 
mass to be above $\simeq 100~{\rm GeV}$.  This forces $m_{\tilde{t}}$ to 
be at least $700~{\rm GeV}$, which in turn leads to $-m_h^2$ larger than 
about $(300~{\rm GeV})^2$ even for $\ln(M_{\rm mess}/m_{\tilde{t}})$ as 
small as a factor of a few (see Eq.~(\ref{eq:corr-Higgs})).  This therefore 
requires a fine-tuning worse than $10\%$.  Although this constraint may 
appear somewhat model-dependent, it in fact applies to rather large classes 
of low-scale supersymmetry breaking theories.  For example, it applies to 
minimal gauge mediation models~\cite{Dine:1994vc}, in which the messenger 
sector (referred to as the supersymmetry-breaking sector here) respects an 
approximate global $SU(5)$ symmetry, i.e. messenger fields fill out complete 
representations of $SU(5)$, and the leading supersymmetry-breaking 
effects are approximately $SU(5)$ symmetric.

It is clear, then, that just making the logarithm smaller does not entirely 
eliminate the fine-tuning.  How much can it help?  To answer this question, 
we make a rough estimate of the minimum value of $M_{\rm mess}$ in the 
following way.  Since the supersymmetry-breaking sector is charged under 
the standard-model gauge group, it contributes not only to the superparticle 
masses but also to the evolution of the standard-model gauge couplings 
above $M_{\rm mess}$.  Suppose now that this sector carries the Dynkin 
index of $\hat{b}$ under $SU(3)_C$, $SU(2)_L$ and $U(1)_Y$ (they are 
universal if this sector respects $SU(5)$).  The requirement that the 
standard-model gauge couplings do not hit the Landau pole below the 
unification scale then gives a constraint on the value of $\hat{b}$. 
For $M_{\rm mess} = O(1\!\sim\!100~{\rm TeV})$, it is $\hat{b} \simlt 5$ 
(for gauge mediation models $\hat{b}$ corresponds to the number of messenger 
fields in ${\bf 5}+{\bf 5}^*$ of $SU(5)$).  The masses of the gauginos 
$\tilde{M}$ and the sfermions $\tilde{m}$ will then be bounded as $\tilde{M} 
\simlt (g^2/16\pi^2) \hat{b} M_{\rm mess}$ and $\tilde{m}^2 \simlt 
(g^2/16\pi^2)^2 C \hat{b} M_{\rm mess}^2$ where $g$ and $C$ represent 
the standard-model gauge coupling and a Casimir factor, because these 
masses are generated as threshold effects at $M_{\rm mess}$ at order 
$g^2$ and $g^4$, respectively.  This gives a bound on the mediation 
scale $M_{\rm mess} \simgt 20~{\rm TeV}$. 

With large mixing for the top squarks, and making $M_{\rm Higgs}$ just 
as large as $114~{\rm GeV}$, one can have $m_{\tilde{t}}$ as small as 
about $700~{\rm GeV}$ (or $500~{\rm GeV}$ if one somehow breaks the 
unified relation of Eq.~(\ref{eq:te-ratio})).  The degree of fine-tuning 
is given by the degree of cancellation between $-m_h^2$ and $|\mu|^2$ 
needed to satisfy Eq.~(\ref{eq:ewsb-cond}).  Since Eq.~(\ref{eq:ewsb-cond}) 
is derived by minimizing the tree-level Higgs potential, and the tree-level 
Higgs quartic coupling of $M_Z^2/4v^2$ is raised to $M_{\rm Higgs}^2/4v^2$ 
by radiative corrections, where $v \equiv (\langle h_u \rangle^2 + 
\langle h_d \rangle^2)^{1/2}$, the real degree of cancellation is 
better measured by Eq.~(\ref{eq:ewsb-cond}) with $M_Z^2$ replaced by 
$M_{\rm Higgs}^2$, i.e. by the fine-tuning parameter
\begin{equation}
  \hat{\Delta}^{-1} \equiv \frac{M_{\rm Higgs}^2/2}{-m_h^2}.
\label{eq:tuning-para}
\end{equation}
We then find that the fine-tuning can be ameliorated to $\hat{\Delta}^{-1} 
\simeq 5\%$ ($\hat{\Delta}^{-1} \simeq 9\%$ in the absence of 
Eq.~(\ref{eq:te-ratio})) in this scenario.  Note, however, that in a 
theory where supersymmetry breaking is mediated by standard-model gauge 
interactions, the fundamental parameter of the theory is not $m_h^2$ but 
the mass scale of the supersymmetry breaking sector, which is proportional 
to the square root of $m_h^2$.  Therefore, if we define the fine-tuning 
parameter as the sensitivity of the weak scale to the fundamental parameters 
of the theory~\cite{Barbieri:1987fn}, it is instead given by
\begin{equation}
  \Delta'^{-1} = \frac{1}{2} \frac{M_{\rm Higgs}^2/2}{-m_h^2} 
    \simlt 3\%,
\label{eq:tuning-para-2}
\end{equation}
($\hat{\Delta}^{-1} \simlt 5\%$ in the absence of Eq.~(\ref{eq:te-ratio})).%
\footnote{In fact, the fine-tuning parameter defined in 
Eq.~(\ref{eq:tuning-para-2}) overestimates the required amount of the tuning 
because it also takes into account generic sensitivity of the weak scale 
to the fundamental parameters~\cite{Anderson:1994dz}.  Nevertheless, we 
will use this parameter to compare the relative amount of tunings required 
in different theories, since it is computed relatively easily and 
unambiguously.  We will discuss this issue in somewhat more detail later.}
Although this level of fine-tuning may not be a real disaster for the 
theory, it is certainly uncomfortable.  Moreover, the origin of the 
large top-squark mixing, which requires a large scalar trilinear 
coupling ($A$ term) of order a few TeV, is not entirely clear.%
\footnote{A possible way to realize this scenario is to consider theories 
in supersymmetric warped space with the Higgs fields propagating in the bulk. 
Supersymmetry is broken on the TeV (infrared) brane while matter fields are 
localized on the Planck (ultraviolet) brane, on which the Yukawa couplings 
are located.  Then, through the TeV-brane coupling between the up-type Higgs 
doublet $H_u$ and the supersymmetry breaking field $Z$ ($\langle Z \rangle 
= \theta^2 F_Z \neq 0$) of the form $\delta(y-\pi R) \int\!d^4\theta 
(Z+Z^\dagger)H_u^\dagger H_u$, the required large top-squark mixing (the large 
trilinear scalar interaction, $A$ term) is generated.  In this theory, however, 
the operators leading to tree-level Higgs soft masses, such as $\delta(y-\pi R) 
\int\!d^4\theta Z^\dagger Z H_u^\dagger H_u$, must be suppressed somehow.}
The origin of the structure of the Higgs sector, e.g. the origin of the 
supersymmetric Higgs mass term ($\mu$ term), is also a mystery, and only 
more so for small values of $M_{\rm mess}$.  In the next section we present 
a scenario which significantly reduces the fine-tuning and avoids these 
difficulties.

To summarize, we have seen that conventional theories of supersymmetry 
breaking, considered widely in the literature, require fine-tuning worse than 
about $3\%$ to reproduce the correct scale of electroweak symmetry breaking. 
Here we emphasize that the fine-tuning discussed in this section becomes even 
severer as the experimental constraints on the Higgs-boson and superparticle 
masses become tighter.  In particular, an increase of the Higgs boson 
mass of a few GeV will push up the value of $m_{\tilde{t}}$ by a large 
amount (because the sensitivity of $M_{\rm Higgs}$ to $m_{\tilde{t}}$ 
is logarithmic) and make the degree of fine-tuning significantly worse. 
The theory we present in the subsequent sections is stable against this 
change, i.e. the theory does not suffer from a severe fine-tuning even 
if the value of $M_{\rm Higgs}$ is increased, as long as $M_{\rm Higgs}$ 
is not much larger than $130~{\rm GeV}$.

\section{Basic Structure of the Theory}
\label{sec:tuningless}

In this section we present a set of general ingredients for constructing 
a theory without severe fine-tuning.  The ideas introduced here will be 
used in constructing an explicit, complete theory in the next section. 

\subsection{Basic elements}
\label{subsec:basic-els}

The discussion from the previous section leads us to demand the 
following properties:
\begin{itemize}
\item[(1)]
We want the scale $M_{\rm mess}$ to be low: we here imagine $M_{\rm mess}$ 
to be of $O(10\!\sim\!100~{\rm TeV})$. 
\item[(2)]
We need an additional source of the physical Higgs-boson mass other than 
those from the $SU(2)_L$ and $U(1)_Y$ $D$-terms and loops of top quark 
and squarks. 
\item[(3)]
We want the superparticle spectrum to be different than that arising 
from an ``$SU(5)$ symmetric'' supersymmetry-breaking sector.  In particular, 
we do not want a large mass hierarchy between colored and non-colored 
superparticles as is the case in Eq.~(\ref{eq:te-ratio}). 
\end{itemize}
To construct a fully realistic and attractive theory with low-energy 
supersymmetry, we want to satisfy these requirements without introducing 
phenomenological problems, and without destroying the successes of the 
SSM.  Specifically, 
\begin{itemize}
\item[(a)]
We want to preserve the successful MSSM prediction associated with gauge 
coupling unification.
\item[(b)]
We do not want the supersymmetric flavor problem to be reintroduced: 
here we require flavor universality for the squark and slepton masses. 
\item[(c)]
We want to have a dark matter candidate in the theory.  In particular, 
we want the candidate to have generic weak-scale cross sections so 
that it naturally provides the observed dark-matter energy density 
as a thermal relic left from the early universe.
\end{itemize}
In addition, we also require the absence of dangerous dimension four 
or five proton decay, and a successful implementation of the see-saw 
mechanism for generating small neutrino masses. 

Satisfying all of requirements~(1)--(3) and (a)--(c) is clearly not 
an easy task.  For example, condition~(3) apparently requires that 
the supersymmetry breaking sector does not respect $SU(5)$.  Such 
a sector generically affects the gauge coupling prediction of the MSSM, 
leading to contradiction with (a).  Condition~(1) makes the lightest 
supersymmetric particle the gravitino, whose mass is expected to be 
of order $M_{\rm mess}^2/M_{\rm Pl} \simeq (0.1\!\sim\!10)~{\rm eV}$, 
and then we lose the lightest neutralino as a dark matter candidate. 
Below we will consider a theory that reconciles these seemingly 
contradictory requirements. 

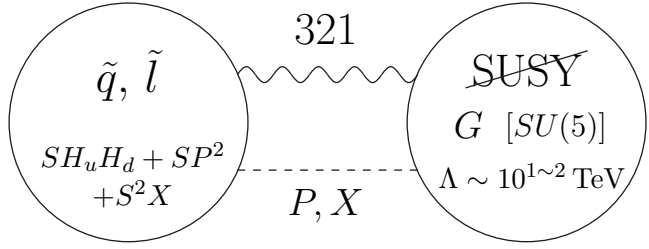
\begin{figure}[t]
\begin{center} 
\begin{picture}(350,90)(0,0)
  \Photon(141,68)(209,68){3}{5}  \Text(175,80)[b]{\Large 321}
  \DashLine(141,32)(209,32){3}   \Text(175,25)[t]{\large $P,X$}
  \CArc(100,50)(45,0,360)
  \Text(100,68)[]{\Large $\tilde{q},\,\tilde{l}$}
  \Text(102,37)[]{\small $SH_uH_d+SP^2$}
  \Text(102,23)[]{\small $+S^2X$}
  \CArc(250,50)(45,0,360)
  \Text(251,70)[]{\Large SUSY}  \Line(227,63)(273,77)
  \Text(229,49)[]{\large $G$} \Text(262,48)[]{[$SU(5)$]}
  \Text(253,30)[]{\small $\Lambda \sim 10^{1 \sim 2}\,{\rm TeV}$}
\end{picture}
\caption{The basic structure of our framework.}
\label{fig:structure}
\end{center}
\end{figure}

The requirements above lead us to the following set of ingredients 
for our theory:
\begin{itemize}
\item[(i)]
We assume that the fundamental scale of supersymmetry breaking is low and 
close to TeV: $M_{\rm mess} = O(10\!\sim\!100~{\rm TeV})$.  This implies 
that we have a sector that induces dynamical supersymmetry breaking (DSB) 
at a scale $\Lambda \sim M_{\rm mess}$.  The DSB sector is charged under 
the standard-model $SU(3)_C \times SU(2)_L \times U(1)_Y$ (321) gauge 
group and leads to 321 gaugino masses of order $(g^2 \hat{b}/16 \pi^2) 
M_{\rm mess}$.  The squark and slepton masses are generated through 
321 gauge interactions and thus are flavor universal. 
\item[(ii)]
The DSB sector possesses an approximate global $SU(5)$ symmetry, whose 
$SU(3) \times SU(2) \times U(1)$ subgroup is weakly gauged and identified 
as the standard model gauge group, 321.  This $SU(5)$ symmetry is then 
broken to the 321 subgroup at the scale $\Lambda$.  Since the global $SU(5)$ 
is broken at $\Lambda \sim M_{\rm mess}$, the three gaugino masses are 
completely independent and the scalar masses do not obey ``unified 
relations'' such as Eq.~(\ref{eq:te-ratio}).  Nevertheless, this broken 
$SU(5)$ symmetry is enough to ensure that the contribution of the DSB 
sector to the 321 gauge coupling evolution is universal at energies larger 
than $\Lambda = O(10\!\sim\!100~{\rm TeV})$, so that the MSSM prediction 
for gauge coupling unification is preserved.  This class of theories was 
first constructed in~\cite{Nomura:2004is}.
\item[(iii)]
We assume that an additional contribution to the physical Higgs-boson mass 
arises from the superpotential coupling $\lambda S H_u H_d$, where $S$ is 
a singlet chiral superfield and $H_u$ and $H_d$ are the two Higgs doublets 
of the MSSM.  While it is possible that some of these fields are composite 
states and their interactions become non-perturbative at low energies, in 
this paper we mainly concentrate on the case where all these fields are 
elementary up to a scale close to the 4D Planck scale. 
\item[(iv)]
Supersymmetry breaking is mediated to the Higgs sector through singlet 
chiral superfields that directly interact both with the DSB and the Higgs 
sectors.  There are many possible variations for the singlet sector.  In 
this paper we mainly consider two classes of singlet fields, which we 
collectively call $X$ and $P$.  The $X$ field couples to $S$ through 
a superpotential term of the form $S^2 X$, while the $P$ field couples 
through $S P^2$.  Through interactions with the DSB sector, $X$ receives 
a VEV and $P$ receives supersymmetry breaking masses.  These in turn 
generate supersymmetric and supersymmetry-breaking terms of order TeV 
in the Higgs sector, generating VEVs for $S$, $H_u$ and $H_d$ of the 
right size.
\item[(v)]
In the explicit model we consider later, we also introduce a singlet field 
$P'$ (or a set of singlets) with exactly the same property as $P$ except 
that it does not directly interact with the DSB sector.  The general couplings 
of the $P$ and $P'$ fields to $S$ then take the form $SP^2 + SPP' + SP'^2$. 
The theory thus has a $Z_2$ discrete symmetry under which the $P$ and 
$P'$ fields are odd while the other fields are even.  If this symmetry 
is unbroken, the lightest member of the $P$ and $P'$ multiplets is a stable 
dark matter candidate.  We call the fields $P$ and $P'$ {\it pedestrian 
fields}, and the $Z_2$ parity acting on these fields {\it pedestrian parity}.
\end{itemize}
A schematic depiction of this framework is given in Fig.~\ref{fig:structure}. 

At first sight, it may seem that constructing a theory possessing all these 
ingredients must require some extremely complicated model building.  However, 
in section~\ref{sec:theory} these features will be incorporated in an 
explicit model in a relatively simple way.  Before constructing this theory, 
we first study the Higgs sector in more detail.

\subsection{Higgs sector}
\label{subsec:new-higgs}

The Higgs sector of our theory contains, in general, a singlet chiral 
superfield $S$, the two Higgs doublets of the MSSM, $H_u$ and $H_d$, and 
a set of singlet fields, $X$, $P$ and $P'$.  To illustrate the basic dynamics 
of the Higgs sector, here we mainly consider only a single pedestrian field 
$P$ that directly interacts with the DSB sector.  The superpotential of the 
Higgs sector then contains the terms
\begin{equation}
  W_H = \lambda S H_u H_d + \frac{\eta}{2} S P^2,
\label{eq:Higgs-superpot}
\end{equation}
where $\lambda$ and $\eta$ are coupling constants.  Since the pedestrian 
field $P$ directly interacts with the DSB sector, it feels the effects 
of supersymmetry breaking through operators of the form
\begin{equation}
  {\cal L} \sim \int\!d^4\theta\, 
    \Bigl( (\hat{Z}^\dagger P^2 + \hat{Z} P^{\dagger 2})
    + \hat{Z}^\dagger \hat{Z} P^\dagger P
    + \hat{Z}^\dagger \hat{Z} (P^2+P^{\dagger 2}) 
    + (\hat{Z}+\hat{Z}^\dagger) P^\dagger P \Bigr),
\label{eq:Higgs-Z}
\end{equation}
where $\hat{Z}$ is a supersymmetry-breaking spurion field, $\langle \hat{Z} 
\rangle = \theta^2 \hat{F}_Z$ with $\hat{F}_Z = O(\Lambda)$, and we have 
omitted the coefficients of the operators.  This generates an effective 
supersymmetric mass term for the $P$ field
\begin{equation}
  W_{{\rm eff},P} = \frac{M_P}{2} P^2,
\label{eq:Pmass-super}
\end{equation}
as well as soft supersymmetry breaking masses 
\begin{equation}
  {\cal L}_{{\rm soft},P} = -m_P^2 |P|^2 - \biggl( \frac{M_P B_P}{2} P^2 
    + \frac{\eta A_\eta}{2} SP^2 + {\rm h.c.} \biggr),
\label{eq:Higgs-soft}
\end{equation}
where we have used the same symbol for a chiral superfield and its 
scalar component.  The parameters $M_P$, $B_P$ and $A_\eta$ are of order 
$c \Lambda$, and $m_P^2$ is of order $(c \Lambda)^2$, where $c$ is a coefficient 
of order $(\hat{b}/16\pi^2)$, so that we naturally expect $|M_P|, |B_P|, 
|A_\eta|, |m_P^2|^{1/2} = O(1\!\sim\!10~{\rm TeV})$.  The Lagrangian given 
by Eqs.~(\ref{eq:Higgs-superpot},~\ref{eq:Pmass-super},~\ref{eq:Higgs-soft}) 
defines our minimal Higgs sector at tree level. 

\begin{figure}[t]
\begin{center} 
\begin{picture}(500,120)(5,110)
  \Text(70,130)[t]{\large (a)}
  \Line(69,150)(69,170) \Line(71,150)(71,170) \Text(75,150)[l]{$F_S$}
  \Line(70,161)(67,158) \Line(70,161)(73,158)
  \DashCArc(70,190)(20,90,270){3} \Text(44,190)[r]{$P$}
  \Line(50,189)(47,192) \Line(50,189)(53,192)
  \DashCArc(70,190)(20,270,90){3} \Text(96,190)[l]{$P$}
  \Line(90,189)(87,192) \Line(90,189)(93,192)
  \Line(67,213)(73,207) \Line(67,207)(73,213)
  \Text(190,130)[t]{\large (b)}
  \DashLine(190,150)(190,170){3} \Text(195,150)[l]{$S$}
  \Line(190,161)(187,158) \Line(190,161)(193,158)
  \DashCArc(190,190)(20,90,270){3} \Text(164,190)[r]{$P$}
  \Line(170,189)(167,192) \Line(170,189)(173,192)
  \DashCArc(190,190)(20,270,90){3} \Text(216,190)[l]{$P$}
  \Line(210,189)(207,192) \Line(210,189)(213,192)
  \Line(187,213)(193,207) \Line(187,207)(193,213)
  \Text(310,130)[t]{\large (c)}
  \DashLine(310,150)(310,170){3} \Text(315,150)[l]{$S$}
  \Line(310,161)(307,158) \Line(310,161)(313,158)
  \DashCArc(310,190)(20,90,270){3} \Text(284,190)[r]{$P$}
  \Line(290,189)(287,192) \Line(290,189)(293,192)
  \DashCArc(310,190)(20,270,90){3} \Text(336,190)[l]{$P$}
  \Line(330,189)(327,192) \Line(330,189)(333,192)
  \Line(309,210)(309,230) \Line(311,210)(311,230) \Text(315,230)[l]{$F_S$}
  \Line(310,221)(307,218) \Line(310,221)(313,218) 
  \Text(430,130)[t]{\large (d)}
  \DashLine(430,150)(430,170){3} \Text(435,150)[l]{$S$}
  \Line(430,161)(427,158) \Line(430,161)(433,158)
  \DashCArc(430,190)(20,90,270){3} \Text(404,190)[r]{$P$}
  \Line(407,193)(413,187) \Line(407,187)(413,193)
  \Line(415,203)(419,203) \Line(415,203)(415,207) 
  \Line(417,175)(413,175) \Line(417,175)(417,179) 
  \CArc(430,190)(19,270,90) \CArc(430,190)(21,270,90) \Text(456,190)[l]{$F_P$}
  \Line(450,189)(447,192) \Line(450,189)(453,192)
  \DashLine(430,210)(430,230){3} \Text(435,230)[l]{$S$}
  \Line(430,221)(427,218) \Line(430,221)(433,218) 
\end{picture}
\caption{The diagrams inducing (a) $F_S$, (b) $S$, (c) $F_S^\dagger S$, 
 and (d) $|S|^2$ terms in the Lagrangian.  The crosses on internal lines 
 represent insertions of supersymmetry-breaking masses.}
\label{fig:loop-Spot}
\end{center}
\end{figure}
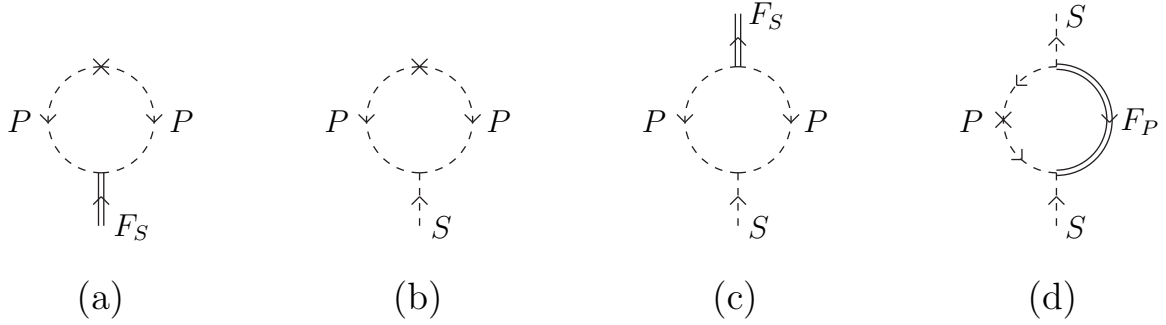
At one loop, the soft supersymmetry breaking terms in Eq.~(\ref{eq:Higgs-soft}) 
induce linear terms in $F_S$ and $S$ in the Lagrangian, through the diagrams 
of Figs.~\ref{fig:loop-Spot}a~and~\ref{fig:loop-Spot}b, where $F_S$ and $S$ 
represent the highest and lowest components of the chiral superfield $S$, 
respectively.  Terms of the form $(F_S^\dagger S + {\rm h.c.})$ and $|S|^2$ 
are also generated through the diagrams of Fig.~\ref{fig:loop-Spot}c and 
Fig.~\ref{fig:loop-Spot}d, respectively.  The term linear in $F_S$ is 
effectively represented by a superpotential term
\begin{equation}
  W_{{\rm eff},S} = L_S^2 S,
\label{eq:Slinear-super}
\end{equation}
which gives terms of the form $(\lambda L_S^{*2} H_u H_d + (\eta/2) L_S^{*2} 
P^2 + {\rm h.c.})$ in the potential.  The other terms in the Lagrangian, 
$(S + {\rm h.c.})$, $(F_S^\dagger S + {\rm h.c.})$ and $|S|^2$, give the 
soft supersymmetry breaking Lagrangian
\begin{equation}
  {\cal L}_{{\rm soft},S} = - \Bigl( L_S^2 C_S S 
    + \lambda A_\lambda S H_u H_d + {\rm h.c.} \Bigr) - m_S^2 |S|^2,
\label{eq:S-soft}
\end{equation}
after eliminating the auxiliary field $F_S$.  Here, $L_S^2$, $L_S^2 C_S$, 
$A_\lambda$ and $m_S^2$ are coefficients.%
\footnote{In the spurion Language, the term in Eq.~(\ref{eq:Slinear-super}) 
arises from the operator of the form $\int\!d^4\theta \{ {\cal D}^2 
(\hat{Z}^\dagger \hat{Z}) S + {\rm h.c.} \}$, where ${\cal D}^2 \equiv 
{\cal D}^\alpha {\cal D}_\alpha$ represents a supercovariant derivative. 
The three terms in Eq.~(\ref{eq:S-soft}), $(S + {\rm h.c.})$, $(F_S^\dagger S 
+ {\rm h.c.})$ and $|S|^2$, arise from $\int\!d^4\theta \{ \hat{Z} {\cal D}^2 
(\hat{Z}^\dagger \hat{Z}) S + {\rm h.c.} \}$, $\int\!d^4\theta \{ \hat{Z} 
S^\dagger S + {\rm h.c.} \}$ and $\int\!d^4\theta \hat{Z}^\dagger \hat{Z} 
S^\dagger S$, respectively.}
Although the diagrams in Fig.~\ref{fig:loop-Spot} are logarithmically 
divergent, the divergences are cut off at the scale $\Lambda$, so that 
the coefficients $L_S^2$, $L_S^2 C_S$, $A_\lambda$ and $m_S^2$ are 
approximately given by 
\begin{eqnarray}
  L_S^2 &\approx& 
    -\frac{\eta}{16\pi^2} M_P^* B_P^* \ln\left(\frac{\Lambda}{|M_P|}\right),
\label{eq:LS2} \\
  L_S^2 C_S &\approx& 
    -\frac{\eta}{16\pi^2} A_\eta M_P^* B_P^* 
        \ln\left(\frac{\Lambda}{|M_P|}\right),
\label{eq:LS2CS} \\
  A_\lambda &\approx&
    -\frac{|\eta|^2}{16\pi^2} A_\eta \ln\left(\frac{\Lambda}{|M_P|}\right),
\label{eq:Alambda} \\
  m_S^2 &\approx& 
    -\frac{|\eta|^2}{8\pi^2} m_P^2 \ln\left(\frac{\Lambda}{|M_P|}\right),
\label{eq:mS2}
\end{eqnarray}
where the logarithm $\ln(\Lambda/|M_P|)$ is expected to be of $O(1)$. 
Altogether, our minimal Higgs sector is effectively given by the superpotentials 
of Eqs.~(\ref{eq:Higgs-superpot},~\ref{eq:Pmass-super},~\ref{eq:Slinear-super}) 
and the supersymmetry breaking Lagrangian of Eqs.~(\ref{eq:Higgs-soft},~%
\ref{eq:S-soft}):%
\footnote{There are also terms in the Lagrangian induced by 
finite corrections, such as $S^\dagger H_u H_d$, $S^\dagger S^2$, 
$S^\dagger P^2$ and $S^\dagger P^\dagger P$, but they are generically 
suppressed and do not affect our analysis significantly.}
\begin{eqnarray}
  W &=& W_H + W_{{\rm eff},P} + W_{{\rm eff},S},
\label{eq:Higgs-W} \\
  {\cal L}_{\rm soft} &=& {\cal L}_{{\rm soft},P} 
    + {\cal L}_{{\rm soft},S} + {\cal L}_{{\rm soft},H}.
\label{eq:Higgs-L}
\end{eqnarray}
Here, we have added in Eq.~(\ref{eq:Higgs-L}) soft supersymmetry-breaking 
mass terms for $H_u$ and $H_d$:
\begin{equation}
  {\cal L}_{{\rm soft},H} = - m_{H_u}^2 |H_u|^2 - m_{H_d}^2 |H_d|^2,
\label{eq:H-soft}
\end{equation}
which arise from 321 gauge loops, with $H_u$ also receiving a contribution 
from the top-squark loop given by Eq.~(\ref{eq:corr-Higgs}).  Note 
that it is possible to modify the minimal Higgs sector by introducing 
additional terms in the tree-level superpotential $W_H$ given in 
Eq.~(\ref{eq:Higgs-superpot}).  We will discuss such a modification 
later in this section.

We now study the vacuum of our Higgs sector.  The VEVs for the fields, 
$H_u$, $H_d$, $S$ and $P$, are given by minimizing the potential derived 
from Eqs.~(\ref{eq:Higgs-W},~\ref{eq:Higgs-L}).  We want to study 
whether our Higgs sector has a phenomenologically acceptable vacuum 
with $\langle S \rangle \neq 0$, $\langle H_u \rangle \neq 0$ and 
$\langle H_d \rangle \neq 0$, and with all physical excitations heavy 
enough to evade experimental constraints.  We are interested in the 
following parameter regions:
\begin{itemize}
\item
The coupling $\lambda$ in $W_H$ should be relatively large so that it can give 
a sizable contribution to the physical Higgs-boson mass, $M_{\rm Higgs}$. 
Specifically, we consider the region where $\lambda \simeq (0.6\!\sim\!0.8)$. 
These values for $\lambda$ are consistent with the requirement that the 
couplings in the Higgs sector do not hit the Landau pole below the unification 
scale $\simeq 10^{16}~{\rm GeV}$.  Note that because the DSB sector is 
charged under the 321 gauge group, high energy values for the 321 gauge 
couplings in our theories are larger than those of the MSSM, which 
allows relatively larger values for $\lambda$ at low energies when evolved 
down from a high scale. (A detailed analysis of the evolution of the 
couplings will be given in section~\ref{subsec:para-Higgs}.)
\item
We consider the region where $\tan\beta \equiv \langle H_u \rangle/\langle 
H_d \rangle$ is not large, say $\tan\beta \simlt 3$.  This is because the 
contribution of $\lambda$ to the square of the physical Higgs-boson mass 
is proportional to $\lambda^2 v^2 \sin^2 2\beta$, which is sizable only 
when $\tan\beta$ is not so large.  Here, $v \equiv (\langle H_u \rangle^2 
+ \langle H_d \rangle^2)^{1/2}$.
\item
We assume that the vacuum preserves pedestrian parity (or $P$ 
parity), under which the $P$ field is odd and the other fields are 
even.  In the context of the present analysis, this is equivalent 
to $\langle P \rangle = 0$. 
\end{itemize}
In a given vacuum, which could be a local minimum of the potential, the 
degree of fine-tuning can be parameterized by the quantity
\begin{equation}
  \Delta^{-1} \equiv \min_i \left| \frac{a_i}{M_Z^2} 
    \frac{\partial M_Z^2}{\partial a_i} \right|^{-1},
\label{eq:ft-parameter}
\end{equation}
which measures the sensitivity of the weak scale to a change of the 
fundamental parameters $a_i$ of the theory~\cite{Barbieri:1987fn}.  This 
quantity reduces to $\hat{\Delta}^{-1}$ in Eq.~(\ref{eq:tuning-para}) 
for the case of the MSSM with $a_i = m_h^2$, and to $\Delta'^{-1}$ of 
Eq.~(\ref{eq:tuning-para-2}) for $a_i \approx \Lambda \propto |m_h^2|^{1/2}$. 
The parameter $\Delta^{-1}$, however, could overestimate the required 
amount of fine-tuning in certain cases.  For example, suppose that $M_Z$ 
itself is the fundamental parameter of the theory, $a_i = M_Z$; then 
Eq.~(\ref{eq:ft-parameter}) gives $\Delta^{-1} = 50\%$ despite the fact 
that the theory is not tuned at all.  This is because $\Delta^{-1}$ also 
takes into account generic sensitivities of $M_Z^2$ to $a_i$, in addition 
to the actual amount of fine-tuning~\cite{Anderson:1994dz}.  To correct 
this, we consider a slightly modified parameter 
\begin{equation}
  \tilde{\Delta}^{-1} \equiv \min_i \left| \eta_i \frac{a_i}{M_Z^2} 
    \frac{\partial M_Z^2}{\partial a_i} \right|^{-1},
\label{eq:ft-parameter-mod}
\end{equation}
where $\eta_i$ are parameters introduced to eliminate generic sensitivities 
of $M_Z^2$ to $a_i$: if $M_Z \propto a_i^n$ in some generic parameter 
region we take $\eta_i = 1/n$.  A difficulty associated with this parameter 
is that it is not easy to estimate $\eta_i$ reliably.  We thus consider 
both $\Delta^{-1}$ and $\tilde{\Delta}^{-1}$ when we perform a detailed 
analysis of electroweak symmetry breaking later.

Since we are looking for a theory which is not severely fine-tuned, we demand 
that the theory has a parameter region that gives $\Delta^{-1} \simgt O(0.1)$ 
at the minimum of the potential.  We consider the minimal Higgs sector defined 
by Eqs.~(\ref{eq:Higgs-superpot},~\ref{eq:Pmass-super},~\ref{eq:Higgs-soft}) 
as well as straightforward modifications obtained by adding terms to 
$W_H$ in Eq.~(\ref{eq:Higgs-superpot}). 

We first observe that the properties of our Higgs potential significantly 
depend on the sign of $m_S^2$.  Let us first consider the case with $m_S^2 
< 0$, which corresponds to $m_P^2 > 0$.  In this case the Higgs potential has 
an unstable direction --- for $H_u = H_d = P =0$, there is a direction in the 
complex $S$ plane in which the potential is not bounded from below, i.e. $V 
\rightarrow -\infty$ for $|S| \rightarrow +\infty$.  This gives a large VEV 
of $S$, at least of order $\Lambda$, and is phenomenologically unacceptable. 
We thus have to introduce a stabilizing term in this case, and the simplest 
possibility is to add a term $(\kappa/3)S^3$ to $W_H$, where $\kappa$ is a 
dimensionless parameter (this also requires the addition of $-(\kappa A_\kappa 
S^3/3 + {\rm h.c.})$ to ${\cal L}_{{\rm soft},S}$ in Eq.~(\ref{eq:S-soft}) 
with $A_\kappa = 3 A_\lambda$).  The theory, then, is essentially the 
next-to-minimal supersymmetric standard model (NMSSM)~\cite{Nilles:1982dy}. 
An important point is that to have a relatively large value for $\lambda$ 
at low energies, $\kappa$ must be small, $\kappa \simlt 0.3$ ($0.4$) for 
$\lambda \simgt 0.7$ ($0.6$).  This is because the RG equation for $\lambda$ 
contains a term proportional to $\lambda \kappa^2$, $d\lambda/d\ln\mu = 
\lambda \kappa^2/8\pi^2 + \cdots$, which gives an asymptotically non-free 
contribution to the evolution of $\lambda$.  With this hierarchy between 
$\lambda$ and $\kappa$ (with $\lambda^2/\kappa^2$ larger than a factor 
of a few), it is difficult to find a parameter region that gives a 
phenomenologically acceptable vacuum with only mild fine-tuning.  One 
typically finds either that the VEV of $S$ is hierarchically smaller than 
the electroweak scale or that there is no electroweak symmetry breaking, 
i.e. either $\langle S \rangle \ll v$ or $v = 0$. (For earlier analyses of 
the NMSSM Higgs sector, though not necessarily for $\lambda^2/\kappa^2 \gg 1$, 
see e.g.~\cite{Dine:1993yw}).  This leads to phenomenologically unacceptable 
consequences.  Therefore, here we do not pursue this possibility further 
and focus on the other case, $m_S^2 > 0$. 

The case with $m_S^2 > 0$ arises when the supersymmetry-breaking mass 
squared for $P$ is negative, $m_P^2 < 0$.  This does not necessarily 
contradict the requirement of unbroken $P$ parity, as long as the 
supersymmetric mass for $P$, $|M_P|$, is sufficiently larger than 
$|m_P^2|^{1/2}$.  In this case, the potential does not have an unstable 
direction.  Therefore, we do not necessarily have to add any additional 
term to stabilize the potential --- our Higgs sector could be simply 
given by Eqs.~(\ref{eq:Higgs-superpot},~\ref{eq:Higgs-Z}) at tree level. 
Taking loop effects into account, the Higgs sector is then effectively 
described by the superpotential
\begin{equation}
  W = \lambda S H_u H_d + L_S^2 S,
\label{eq:Higgs2-W}
\end{equation}
and the soft supersymmetry-breaking Lagrangian
\begin{equation}
  {\cal L}_{\rm soft} 
    = - m_{H_u}^2 |H_u|^2 - m_{H_d}^2 |H_d|^2 - m_S^2 |S|^2 
      - \Bigl( \lambda A_\lambda S H_u H_d + L_S^2 C_S S + {\rm h.c.} \Bigr),
\label{eq:Higgs2-L}
\end{equation}
where we have set $P = 0$.  The parameters $L_S^2$, $m_S^2$, $A_\lambda$ 
and $L_S^2 C_S$ are related to the fundamental parameters of the theory 
through Eqs.~(\ref{eq:LS2}--\ref{eq:mS2}).  (For earlier studies of the 
Higgs potential of the form Eqs.~(\ref{eq:Higgs2-W},~\ref{eq:Higgs2-L}), 
see e.g.~\cite{Panagiotakopoulos:2000wp}.)

It is possible that there are additional terms in the Higgs-sector 
superpotential of Eq.~(\ref{eq:Higgs2-W}).  An obvious example is
\begin{equation}
  \delta W = \frac{\kappa}{3} S^3.
\label{eq:S3}
\end{equation}
It is also possible that the Higgs-sector superpotential has the term 
\begin{equation}
  \delta W = \frac{M_S}{2} S^2,
\label{eq:S2}
\end{equation}
where $M_S$ is a parameter of order the weak scale.  A term of this form 
can arise effectively if there is a singlet field $X$ that couples to $S$ 
as $W = S^2 X$ and receives a VEV of order the weak scale through direct 
interactions to the DSB sector (an explicit realization of this will 
be considered in the next section).  The presence of the terms in 
Eqs.~(\ref{eq:S3},~\ref{eq:S2}) can affect the phenomenology of the theory, 
especially electroweak symmetry breaking and the neutralino spectrum.  In 
the rest of this subsection, however, we focus on the simplest superpotential 
of Eq.~(\ref{eq:Higgs2-W}), and discuss its consequences on electroweak 
symmetry breaking.  The terms Eqs.~(\ref{eq:S3},~\ref{eq:S2}) will be 
considered in later sections.

Denoting the neutral components of $H_u$ and $H_d$, i.e. the 
components that will get VEVs, as $H_u^0$ and $H_d^0$, and setting 
the charged components to be zero, the scalar potential derived from 
Eqs.~(\ref{eq:Higgs2-W},~\ref{eq:Higgs2-L}) is 
\begin{eqnarray}
  V = V_F + V_D + V_{\rm soft},
\label{eq:Higgs-Pot}
\end{eqnarray}
where $V_F$, $V_D$ and $V_{\rm soft}$ are given by
\begin{eqnarray}
  V_F &=& |\lambda H_u^0 H_d^0 - L_S^2|^2 
    + |\lambda|^2 |S|^2 (|H_u^0|^2+|H_d^0|^2),
\\
  V_D &=& \frac{g^2+g'^2}{8} (|H_u^0|^2 - |H_d^0|^2)^2,
\\
  V_{\rm soft} &=& m_{H_u}^2 |H_u^0|^2 + m_{H_d}^2 |H_d^0|^2 + m_S^2 |S|^2 
    + \Bigl( -\lambda A_\lambda S H_u^0 H_d^0 + L_S^2 C_S S + {\rm h.c.} \Bigr).
\end{eqnarray}
Here, $V_D$ arises from the $SU(2)_L$ and $U(1)_Y$ $D$-terms, and $g$ and 
$g'$ represent the $SU(2)_L$ and $U(1)_Y$ gauge couplings. 

An important feature of the Higgs potential of Eq.~(\ref{eq:Higgs-Pot}) is 
that, unlike the case where the VEV of $S$ is stabilized by a small coupling 
$\kappa$, the VEVs of $S$ and the Higgs doublets can be determined essentially 
by independent conditions.  To see this, let us look at the derivative of 
the potential in terms of the fields.  Assuming that the parameters in the 
potential are all real, for simplicity, we find
\begin{eqnarray}
  \frac{\partial V}{\partial H_u^{0\dagger}} 
  &=& \lambda H_d^{0\dagger} (\lambda H_u^0 H_d^0 - L_S^2) 
    + \lambda^2 H_u^0 |S|^2 \nonumber \\
  && + \frac{g^2+g'^2}{4} H_u^0 (|H_u^0|^2 - |H_d^0|^2) + m_{H_u}^2 H_u^0 
    - \lambda A_\lambda S^\dagger H_d^{0\dagger},
\label{eq:V-der-Hu} \\
  \frac{\partial V}{\partial H_d^{0\dagger}} 
  &=& \lambda H_u^{0\dagger} (\lambda H_u^0 H_d^0 - L_S^2) 
    + \lambda^2 H_d^0 |S|^2 \nonumber \\
  && - \frac{g^2+g'^2}{4} H_d^0 (|H_u^0|^2 - |H_d^0|^2) + m_{H_d}^2 H_d^0 
    - \lambda A_\lambda S^\dagger H_u^{0\dagger},
\label{eq:V-der-Hd} \\
  \frac{\partial V}{\partial S^\dagger} 
  &=& \lambda^2 S (|H_u^0|^2+|H_d^0|^2) + m_S^2 S 
    - \lambda A_\lambda H_u^{0\dagger} H_d^{0\dagger} + L_S^2 C_S.
\label{eq:V-der-S}
\end{eqnarray}
We want our vacuum to be at $v^2 \equiv |\langle H_u^0 \rangle|^2 + |\langle 
H_d^0 \rangle|^2 \simeq (174~{\rm GeV})^2$, $\tan\beta \equiv |\langle H_u^0 
\rangle/\langle H_d^0 \rangle| \simlt 3$ and $|\mu_{\rm eff}| \equiv \lambda 
|\langle S \rangle| \simgt 100~{\rm GeV}$, and these values must be given 
as a solution of $\partial V/\partial H_u^{0\dagger} = \partial V/\partial 
H_d^{0\dagger} = \partial V/\partial S^\dagger = 0$. 

Suppose now that the supersymmetry-breaking masses associated with $S$, 
specifically $|L_S^2 C_S|^{1/3}$ and $(m_S^2)^{1/2}$, are somewhat larger 
than $v$, with sizes $\approx (400\!\sim\!800)~{\rm GeV}$. 
In this case, $\partial V/\partial S^\dagger$ is dominated by the second 
and the last terms in the right-hand-side of Eq.~(\ref{eq:V-der-S}). 
This implies that the VEV of $S$ is essentially determined by the balance 
between these two terms in $\partial V/\partial S^\dagger = 0$, i.e. the 
balance between the linear and quadratic terms in $S$ in the potential, 
irrespective of the dynamics determining the VEVs of $H_u^0$ and $H_d^0$:
\begin{equation}
  \langle S \rangle \simeq - \frac{L_S^2 C_S}{m_S^2}.
\end{equation}
Since the mass squared of $S$ is given by $m_S^2 \gg v^2$, the VEV of 
$S$ can be regarded as essentially fixed when one considers the minimization 
with respect to the VEVs of $H_u^0$ and $H_d^0$ using $\partial V/\partial 
H_u^{0\dagger} = \partial V/\partial H_d^{0\dagger} = 0$.  Assuming real 
VEVs, we obtain from Eq.~(\ref{eq:V-der-Hu},~\ref{eq:V-der-Hd})
\begin{eqnarray}
  \frac{g^2+g'^2}{4}v^2 
    &=& \frac{m_{H_d}^2-\tan^2\!\beta\, m_{H_u}^2}{\tan^2\!\beta-1}
    - \lambda^2 S^2,
\\
  (m_{H_u}^2 + m_{H_d}^2 + 2 \lambda^2 S^2)\frac{\sin 2\beta}{2}
    &=& \lambda(L_S^2 + A_\lambda S - \lambda \sin\beta \cos\beta\, v^2).
\end{eqnarray}
These equations are identical to the MSSM minimization conditions, with 
the $\mu$ and $\mu B$ parameters of the MSSM identified with the effective 
$\mu$ and $\mu B$ parameters defined by $\mu_{\rm eff} \equiv \lambda 
\langle S \rangle$ and $(\mu B)_{\rm eff} \equiv \lambda(L_S^2 + A_\lambda 
\langle S \rangle - \lambda \sin\beta \cos\beta\, v^2)$.  Therefore, we 
find that for $\mu_{\rm eff} \approx (\mu B)_{\rm eff}^{1/2} \approx 
(100\!\sim\!200)~{\rm GeV}$, our Higgs sector produces realistic 
electroweak symmetry breaking without severe fine-tuning.  For smaller 
$|L_S^2 C_S|^{1/3}$ and $(m_S^2)^{1/2}$, the situation is somewhat more 
complicated as the dynamics determining $\langle S \rangle$ and $v$ are 
coupled with each other. 

The desired parameter region, $\mu_{\rm eff} \approx (\mu B)_{\rm eff}^{1/2} 
\simlt v$, corresponds to the region where the combination of parameters 
$|A_\eta M_P B_P/m_P^2|$ is smaller than its ``natural value'' of order 
$c\Lambda$ by one or two orders of magnitude, depending on the size of 
$\eta$ (for $\eta \simgt 1$, $|M_P B_P|$ must also be somewhat suppressed 
compared with $(c \Lambda)^2$).  This may be achieved, for example, by 
simply talking the value of $M_P B_P$ smaller than its ``natural value'' 
$\approx (c\Lambda)^2$.  It is important to notice that this does not 
necessarily lead to a fine-tuning in electroweak symmetry breaking, since 
we have simply chosen some parameters to be small and have not required 
any precise cancellation between independent quantities.  In fact, any 
fractional change of parameters in our Higgs potential leads to a fractional 
change of the weak scale, $M_Z$, of roughly the ``same'' amount, so that the 
fine-tuning parameter $\Delta^{-1}$ defined in Eq.~(\ref{eq:ft-parameter}) 
is not much smaller than one: typically $\Delta^{-1} = O(20\!\sim\!30\%)$.%
\footnote{In the context of an explicit model, one must check that the 
desired values $\mu_{\rm eff} \approx (\mu B)_{\rm eff}^{1/2} \simlt v$ 
are obtained without any hidden fine-tuning.  For example, $M_P B_P \ll 
(c\Lambda)^2$ may not be realized naturally if the DSB sector is truly 
strongly coupled and all the parameters obey naive dimensional analysis. 
In our explicit model given in the next section, we obtain the desired 
values naturally by introducing an additional pedestrian field $P'$ 
that interacts with $S$ but does not interact with the DSB sector.}

We finally emphasize some of the virtues of our Higgs potential, effectively 
described by Eqs.~(\ref{eq:Higgs2-W},~\ref{eq:Higgs2-L}), over the NMSSM 
Higgs potential, which is more commonly considered in the literature.  First 
of all, our Higgs potential leads to realistic electroweak symmetry breaking 
much more easily, as we have seen in this subsection.  This is particularly 
true when the coupling $\lambda$ is taken to be large to push the physical 
Higgs-boson mass larger.  The sensitivity of $M_Z$ to the fundamental 
parameters is much weaker, allowing for reduced fine-tuning.  Moreover, 
our Higgs potential does not have any approximate continuous symmetry that 
leads to an unwanted light state in the spectrum when it is spontaneously 
broken by the non-zero VEVs of $S$, $H_u$ and $H_d$.  The discrete $Z_3$ 
symmetry of the NMSSM potential, which leads to the cosmological domain 
wall problem, is also absent in our potential, as is evident from the 
form of Eqs.~(\ref{eq:Higgs2-W},~\ref{eq:Higgs2-L}).  Finally, it is also 
interesting to note that we do not need a large $S^3$ coupling in the 
superpotential to stabilize the $S$ VEV.  This helps us obtain larger 
values of $\lambda$ at the weak scale, and thus larger values of the 
physical Higgs-boson mass through the contribution from the $S H_u H_d$ 
term in the superpotential.

\section{Models}
\label{sec:theory}

In this section we explicitly construct a theory accommodating all the 
features described in the previous section.  We first explain how the 
picture described in section~\ref{subsec:basic-els} leads us to consider 
a certain class of theories --- theories in 5D warped spacetime with 
supersymmetry broken at the infrared (IR) brane and the bulk unified 
symmetry broken both on the ultraviolet (UV) brane and on the IR brane. 
We then describe explicit models accommodating not only the basic structure 
of section~\ref{subsec:basic-els} but also the structure of the Higgs 
sector discussed in section~\ref{subsec:new-higgs}.  Electroweak 
symmetry breaking in these models will be studied in the next section.

\subsection{Supersymmetric unification in warped space}
\label{subsec:sup-unif}

As discussed in section~\ref{subsec:basic-els}, the basic ingredients 
of our theory are as depicted in Fig.~\ref{fig:structure}. In particular, 
there must be a sector, the DSB sector, that dynamically breaks 
supersymmetry at the scale near TeV.  We denote the gauge group of 
this sector as $G$, and its gauge coupling and the size (the number 
of ``colors'') as $\tilde{g}$ and $\tilde{N}$, respectively.  The DSB 
sector is charged under the standard-model 321 gauge interaction.  Once 
supersymmetry is broken, the breaking is directly transmitted to the 321 
gauginos, giving them masses of order TeV.  The squarks and sleptons feel 
supersymmetry breaking only through the 321 gauge loops, so that they 
receive flavor universal masses at loop level.  At first sight, it may 
seem that this type of scenario does not allow for a calculable theory 
because it necessarily involves strong dynamics near the TeV scale, 
$\Lambda \approx (10\!\sim\!100)~{\rm TeV}$.  However, if the DSB sector 
satisfies certain special properties we can formulate a calculable theory 
using a ``dual'' higher dimensional description of the theory. 

Suppose that the gauge group $G$, responsible for dynamical supersymmetry 
breaking, has a large 't~Hooft coupling and a large number of colors, 
i.e. $\tilde{\kappa} \equiv \tilde{g}^2 \tilde{N}/16\pi^2 \gg 1$ and 
$\tilde{N} \gg 1$.  We also assume that the coupling $\tilde{\kappa}$ 
is almost constant above the TeV scale.  In this case the AdS/CFT 
correspondence~\cite{Maldacena:1997re,Arkani-Hamed:2000ds} suggests 
that we can formulate this theory in 5D anti-de~Sitter (AdS) spacetime 
truncated by two branes.  The resulting 5D theory appears as a supersymmetric 
theory on warped space ($0 \leq y \leq \pi R$) with the metric given by
\begin{equation}
  ds^2 = e^{-2ky} \eta_{\mu\nu} dx^\mu dx^\nu + dy^2,
\label{eq:metric}
\end{equation}
where $y$ is the coordinate for the extra dimension and $k$ denotes the 
inverse curvature radius of the AdS space~\cite{Randall:1999ee}.  The two 
branes are located at $y=0$ and $y=\pi R$, which are called the UV brane 
(or the Planck brane) and the IR brane (or the TeV brane), respectively. 
The scales are chosen such that the scales on the UV and IR branes 
are roughly the 4D Planck scale and the scale $\Lambda$, respectively: 
$k \sim M_5 \sim M_* \sim M_{\rm Pl}$ and $kR \sim 10$.  Here, $M_5$ is 
the 5D Planck scale, and $M_*$ the 5D cutoff scale, which is taken to 
be somewhat (typically a factor of a few) larger than $k$.  The 4D Planck 
scale is given by $M_{\rm Pl}^2 \simeq M_5^3/k$ and the scale on the 
IR brane is defined as $k' \equiv k e^{-\pi kR} \sim \Lambda$.%
\footnote{The description of the various scales here is quite rough, and does 
not discriminate two scales that differ by one or two orders of magnitude. 
A more precise choice of  scales will be made later, where we will 
identify $k$ as the unification scale $\simeq 10^{16}~{\rm GeV}$ and 
choose $k' \simeq (10\!\sim\!100)~{\rm TeV}$.}

The standard model gauge fields propagate in the 5D bulk and quarks and 
leptons are localized to the Planck brane.  Supersymmetry breaking caused 
by the IR dynamics of $G$ then corresponds to supersymmetry breaking localized 
on the TeV brane.  The masses for the 321 gauginos are generated at tree level 
though their interactions on the TeV brane, which in turn generate squark 
and slepton masses at one loop.  This, therefore, corresponds to the class of 
theories considered in~[\ref{Gherghetta:2000qt:X}~--~\ref{Nomura:2004it:X},~%
\ref{Nomura:2004is:X}].  As shown in~\cite{Goldberger:2002pc}, this class 
of theories leaves many of the most attractive features of conventional 
unification intact.  In particular, the successful MSSM prediction 
for the low-energy gauge couplings is preserved, provided that the 5D 
bulk possesses an $SU(5)$ gauge symmetry which is broken at the Planck 
brane and that matter and two Higgs doublets are localized towards the 
Planck brane or have conformally-flat wavefunctions (for earlier work 
see~\cite{Pomarol:2000hp}).  Any physics that uses high scales can be 
accommodated without any obstacle; for example, small neutrino masses can 
be generated naturally through the see-saw mechanism.  Note, however, that 
for certain purposes the theories reveal their higher-dimensional nature at 
a scale not far from a TeV, through the appearance of Kaluza-Klein (KK) 
towers and an $N=2$ supermultiplet structure.  This cuts off divergences 
in supersymmetry-breaking quantities at the KK scale $k'$ and allows 
small values of $M_{\rm mess} \sim k' = O(10\!\sim\!100~{\rm 
TeV})$~\cite{Nomura:2003qb}.  This class of theories, therefore, 
naturally incorporates ingredient~(i) in section~\ref{subsec:basic-els}. 

Here we emphasize that we should not take the view that we have solved 
the hierarchy problem twice by introducing both a warped extra dimension 
and supersymmetry.  Rather, our 5D theory is obtained by requiring 
certain properties on the DSB sector, which is necessarily present in 
any low-energy supersymmetric theory.  For example, we require that the 
parameters $\tilde{\kappa}$ and $\tilde{N}$ in the DSB sector are large, 
the evolution of $\tilde{\kappa}$ is very slow over a wide energy interval 
between $k$ and $k'$, and the IR dynamics of $G$ produces certain gaps 
among the anomalous dimensions of various different $G$-invariant operators. 
These requirements then naturally lead to a supersymmetric warped extra 
dimension in the ``dual'' higher dimensional (5D) description.%
\footnote{Strictly speaking, it is not necessarily guaranteed that the 4D 
theory corresponding to our 5D theory (or the consistent embedding of the 
5D theory into string theory) exists.  Here we do not address the issue 
of constructing a fully UV completed theory, and treat our 5D theory in 
the context of an effective higher-dimensional field theory.}
This viewpoint was particularly emphasized in Ref.~\cite{Nomura:2004zs}, 
which we follow here.

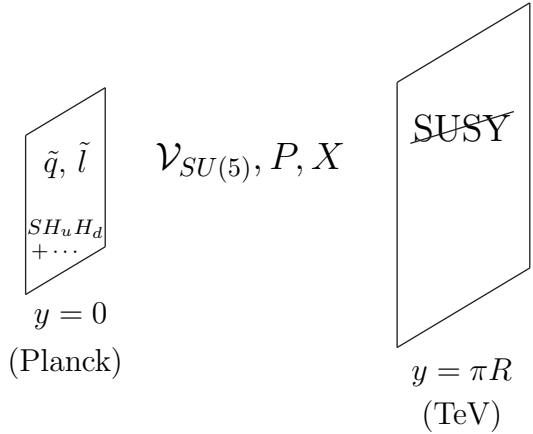
\begin{figure}[t]
\begin{center} 
\begin{picture}(190,160)(0,-35)
  \Text(85,70)[]{\large ${\cal V}_{SU(5)}, P, X$}
  \Line(0,20)(0,80) \Line(30,38)(30,98)
  \Line(0,20)(30,38) \Line(0,80)(30,98)
  \Text(15,71)[]{$\tilde{q},\,\tilde{l}$}
  \Text(15,45)[]{\scriptsize $SH_uH_d$}
  \Text(2,36)[l]{\scriptsize $+\cdots$}
  \Text(17,17)[t]{$y=0$}  \Text(15,0)[t]{(Planck)}
  \Line(140,0)(140,100) \Line(190,30)(190,130)
  \Line(140,0)(190,30) \Line(140,100)(190,130)
  \Text(165,83)[]{\large SUSY}  \Line(145,77)(184,89)
  \Text(165,-4)[t]{$y=\pi R$}  \Text(165,-20)[t]{(TeV)}
\end{picture}
\caption{The 5D picture of our theory.}
\label{fig:5D}
\end{center}
\end{figure}

Let us now look at the group theoretical structure of our theory more 
carefully.  To preserve the successful MSSM prediction for the low-energy 
gauge couplings, the 5D bulk must respect, at least, $SU(5)$, which is broken 
to the 321 subgroup at the Planck brane.  In the 4D picture this corresponds 
to the DSB sector $G$ possessing a global $SU(5)$ symmetry, of which only 
the 321 subgroup is gauged (at least at energies below $\sim k$). This 
is crucial for the $G$ sector not to destroy the successful prediction of 
the MSSM.  In our framework the global $SU(5)$ symmetry in the DSB sector 
should further be broken down to 321 at the scale $\Lambda$, in order for 
the superparticle spectrum not to have $SU(5)$-symmetric features (this is 
ingredient~(ii) from section~\ref{subsec:basic-els}).  In the 5D picture 
this implies that the TeV brane respects only the 321 subgroup of $SU(5)$, 
as can be attained by breaking the bulk $SU(5)$ symmetry to 321 by boundary 
conditions at the TeV brane.  We therefore arrive at the following picture 
for the structure of our 5D theory.  The bulk gauge group is $SU(5)$ and 
it is broken to the 321 subgroup both at the Planck and TeV branes.  The 
MSSM quark and lepton fields are localized on the Planck brane, together 
with the two Higgs doublets $H_u$ and $H_d$ and a singlet field $S$.  The 
schematic picture of this 5D theory is depicted in Fig.~\ref{fig:5D}.

Supersymmetry breaking effects are mediated to the sector localized 
to the Planck brane in essentially two different ways --- through 321 
gauge interactions and through bulk singlet fields.  We introduce a 
singlet field $P$ in the bulk so that it has non-negligible interactions 
both to the Planck and TeV branes.  This is one of the pedestrian fields 
discussed earlier, which couples to $S$ on the Planck brane as $W = SP^2$. 
We also introduce one (or more) pedestrian field $P'$ on the Planck 
brane, which also couples to $S$, as was discussed in ingredient~(v) 
in section~\ref{subsec:basic-els}.  Finally, depending on the specific 
model, we may also introduce singlet fields, which we collectively call 
$X$, that transmit the scale of the TeV brane to the Planck brane and 
generate supersymmetric masses of $S$ and/or $P'$ fields of order the 
weak scale.  Our explicit models will be described in more detail in 
the next two subsections. 

\subsection{Structure of models}
\label{subsec:warped}

We have identified the gauge symmetry structure of our 5D theory --- the 
bulk possesses an $SU(5)$ symmetry, which is broken to the 321 subgroup 
both on the Planck and the TeV branes.  Warped unified models with the 
bulk $SU(5)$ gauge symmetry broken to 321 both at the Planck and TeV branes 
were constructed in Ref.~\cite{Nomura:2004is}.  Here we construct our 
model along the lines of section~\ref{sec:tuningless}, following the 
basic construction of~\cite{Nomura:2004is}.

The theory is formulated in a 5D warped spacetime with the extra dimension 
$y$ compactified on $S^1/Z_2$ ($0 \leq y \leq \pi R$).  The metric is given 
by Eq.~(\ref{eq:metric}), and parameters are chosen as $k \sim M_5 \sim 
M_* \sim M_{\rm Pl}$ and $kR \sim 10$.  We consider a supersymmetric 
$SU(5)$ gauge theory on this gravitational background, with the bulk 
$SU(5)$ symmetry broken by boundary conditions both at $y=0$ and $\pi R$.%
\footnote{Alternatively, the $SU(5)$ breaking could be caused by 
an $SU(5)$-breaking Higgs field(s) localized on a brane(s).  For the 
TeV-brane breaking, the VEV of the brane Higgs field must be close to 
the (local) cutoff scale so that the resulting superparticle spectrum 
does not have a characteristic feature arising from the $SU(5)$ symmetry. 
On the other hand, the Planck-brane breaking can be caused by an 
$SU(5)$-breaking VEV not necessarily close to the cutoff scale, in which 
case the unified scale is identified as the smaller of the VEV and the 
AdS curvature scale $k$.}
Specifically, the 5D gauge multiplet can be decomposed into a 4D $N=1$ 
vector superfield $V(A_\mu, \lambda)$ and a 4D $N=1$ chiral superfield 
$\Sigma(\sigma+iA_5, \lambda')$, where both $V$ and $\Sigma$ are in the 
adjoint representation of $SU(5)$. The boundary conditions for these fields 
are given by
\begin{equation}
  \pmatrix{V \cr \Sigma}(x^\mu,-y) 
  = \pmatrix{ \hat{P} V \hat{P}^{-1} \cr 
             -\hat{P} \Sigma \hat{P}^{-1}}(x^\mu,y), 
\qquad
  \pmatrix{V \cr \Sigma}(x^\mu,-y') 
  = \pmatrix{ \hat{P} V \hat{P}^{-1} \cr 
             -\hat{P} \Sigma \hat{P}^{-1}}(x^\mu,y'), 
\label{eq:bc-g}
\end{equation}
where $y' = y - \pi R$, and $\hat{P}$ is a $5 \times 5$ matrix acting on 
gauge space: $\hat{P} = {\rm diag}(+,+,+,-,-)$.  This reduces the gauge 
symmetry to 321 both at the Planck and TeV branes.  The gauge symmetry 
at low energies is 321.  The zero-mode sector contains not only the 
321 component of $V$, which is the gauge multiplet of the SSM, but 
also the $SU(5)/(SU(3)_C \times SU(2)_L \times U(1)_Y)$ (XY) component 
of $\Sigma$.  The typical mass scale for the KK towers is $\pi k' = 
O(100~{\rm TeV})$, so that the lowest KK excitations of the standard model 
gauge fields and the lightest XY gauge bosons both have masses of order 
$100~{\rm TeV}$. 

Matter and Higgs fields are introduced on the Planck brane.  We introduce 
a standard set of matter chiral superfields $Q({\bf 3}, {\bf 2})_{1/6}$, 
$U({\bf 3}^*, {\bf 1})_{-2/3}$, $D({\bf 3}^*, {\bf 1})_{1/3}$, $L({\bf 1}, 
{\bf 2})_{-1/2}$ and $E({\bf 1}, {\bf 1})_{1}$ for each generation, where 
the numbers represent the transformation properties under 321 with the 
$U(1)_Y$ charges normalized in the conventional way.%
\footnote{In the case of boundary condition breaking, there is a priori 
no reason why the hypercharges for matter fields are quantized with $SU(5)$ 
normalization.  The quantization could be understood either in the case 
of Higgs $SU(5)$ breaking or in a more fundamental theory such as one 
with a larger gauge group or in higher dimensions.  Alternatively, we can 
obtain a partial understanding of the matter quantum numbers by slightly 
delocalizing matter fields in the bulk.  Such a delocalization, in fact, 
may be needed to avoid a dangerous thermal relic of the lightest XY 
state of $\Sigma$, $A_5^{\rm XY}$, in the universe~\cite{Nomura:2005qg}.}
For the Higgs sector, we introduce three chiral superfields $S({\bf 1}, 
{\bf 1})_{0}$, $H_u({\bf 1}, {\bf 2})_{1/2}$ and $H_d({\bf 1}, 
{\bf 2})_{-1/2}$ on the Planck brane.%
\footnote{These fields could alternatively be introduced in the bulk as 
hypermultiplets obeying appropriate boundary conditions.  This reproduces 
essentially the same physics as the brane-field case if we localize the 
zero-mode wavefunctions towards the Planck brane by large bulk hypermultiplet 
masses, $c_S, c_{H_u}, c_{H_d} \gg 1/2$ (for the definition of $c$ parameters 
see Eq.~(\ref{eq:bulk-mass})).}
The Yukawa couplings are then written on the Planck brane:
\begin{equation}
  S_{\rm Yukawa} = \int\!d^4x \int_0^{\pi R}\!\!dy \,\, 
    2 \delta(y) \biggl[ \int\!d^2\theta \left( y_u Q U H_u 
    + y_d Q D H_d + y_e L E H_d \right) + {\rm h.c.} \biggr],
\label{eq:yukawa-321}
\end{equation}
where we have suppressed generation indices.

As discussed before, we introduce a pedestrian field $P$ in the bulk 
and one (or more) pedestrian field $P'$ on the Planck brane.  The field 
$P$ transmits supersymmetry breaking from the TeV brane to the Higgs 
sector on the Planck brane.  Using notation where a bulk hypermultiplet 
is represented by two 4D $N=1$ chiral superfields $\Phi(\phi,\psi)$ and 
$\Phi^c(\phi^c,\psi^c)$ with opposite quantum numbers, the bulk pedestrian 
field $P$ can be written as $\{ P, P^c \}$.  Without loss of generality, 
we choose the boundary conditions for this field as
\begin{equation}
  \pmatrix{P \cr P^c}(x^\mu,-y) 
  = \pmatrix{P \cr -P^c}(x^\mu,y), 
\qquad
  \pmatrix{P \cr P^c}(x^\mu,-y') 
  = \pmatrix{P \cr -P^c}(x^\mu,y').
\label{eq:bc-p}
\end{equation}
A bulk hypermultiplet $\{ \Phi, \Phi^c \}$ can generically have a mass 
term in the bulk, which is written as 
\begin{equation}
  S_{\Phi} = \int\!d^4x \int_0^{\pi R}\!\!dy \, 
    \biggl[ e^{-3k|y|}\! \int\!d^2\theta\, c_\Phi k \Phi \Phi^c 
    + {\rm h.c.} \biggr],
\label{eq:bulk-mass}
\end{equation}
in the basis where the kinetic term is given by $S_{\rm kin} = \int\!d^4x 
\int\!dy\, [e^{-2k|y|} \int\!d^4\theta (\Phi^\dagger \Phi + \Phi^c 
\Phi^{c\dagger}) + \{ e^{-3k|y|} \int\!d^2\theta (\Phi^c \partial_y \Phi 
- \Phi \partial_y \Phi^c)/2 + {\rm h.c.} \}]$~\cite{Marti:2001iw}.  The 
parameter $c_\Phi$ controls the wavefunction profile of the zero mode. 
For $c_\Phi > 1/2$ ($< 1/2$) the wavefunction of a zero mode arising 
from $\Phi$ is localized to the Planck (TeV) brane; for $c_\Phi = 1/2$ 
it is conformally flat.  We choose $c_P \simeq 1/2$ so that the zero mode 
arising from $P$ has a nearly conformally flat wavefunction and that a 
large exponential suppression does not arise when transmitting supersymmetry 
breaking from the TeV brane to the Planck brane. 

The couplings of the pedestrian fields to the Higgs sector are given by
\begin{equation}
  S_{\rm Higgs} = \int\!d^4x \int_0^{\pi R}\!\!dy \,\, 
    2 \delta(y) \biggl[ \int\!d^2\theta \left( \lambda S H_u H_d 
    + \frac{\eta}{2} S P^2 + \eta' S P P' + \frac{h}{2} S P'^2 \right)
    + {\rm h.c.} \biggr].
\label{eq:Higgs-superpot-5D}
\end{equation}
Note that the superpotential of Eqs.~(\ref{eq:yukawa-321})~and~%
(\ref{eq:Higgs-superpot-5D}), as well as the bulk Lagrangian, is 
invariant under a $U(1)_R$ symmetry under which various fields transform 
as $V(0)$, $\Sigma(0)$, $Q(1)$, $U(1)$, $D(1)$, $L(1)$, $E(1)$, $H_u(0)$, 
$H_d(0)$, $S(2)$, $P(0)$, $P^c(2)$ and $P'(0)$ in the normalization 
where the superpotential has a charge of $+2$.  Imposing this symmetry, 
potentially dangerous operators on the Planck brane, such as the ones 
leading to rapid proton decay and a large mass for the Higgs doublets, 
are forbidden.  In particular, all dimension four and five proton decay 
operators are forbidden by the $U(1)_R$ symmetry.  We also introduce 
a discrete $Z_2$ symmetry, under which $P$, $P^c$ and $P'$ are odd and 
all the other fields are even.  This is the pedestrian parity (or $P$ 
parity) discussed in section~\ref{sec:tuningless}.  After supersymmetry 
is broken on the TeV brane, $U(1)_R$ is broken to the $Z_{2,R}$ subgroup, 
which is exactly the $R$ parity of the MSSM.  We assume that the $P$ parity 
remains unbroken even after supersymmetry breaking.  The unbroken $P$ parity 
ensures the stability of the lightest component of the pedestrian fields, 
making it a candidate for the dark matter of the universe (this issue 
will be discussed in section~\ref{subsec:pedestrian-DM}).

With the above configuration of fields, the successes of conventional 
supersymmetric unification are preserved~\cite{Nomura:2004is}.  In 
particular, assuming that tree-level brane-localized kinetic terms 
are small as suggested by naive dimensional analysis (which corresponds 
in the 4D picture to the assumption that the 321 gauge couplings become 
strong at the scale $\approx k$),%
\footnote{This assumption is not needed for the Planck-brane localized 
kinetic terms if the $SU(5)$ breaking on the Planck brane is caused by 
the $SU(5)$-breaking Higgs field with the VEV hierarchically (one or two 
orders of magnitude) smaller than the cutoff scale.  In this case, the 
$SU(5)$-breaking VEV appears in Eq.~(\ref{eq:gc-low}), instead of $k$, 
and is identified as the unification scale.}
the low-energy 321 gauge couplings $g_a$ ($a=1,2,3$) are given by
\begin{equation}
  \pmatrix{1/g_1^2(k') \cr 1/g_2^2(k') \cr 1/g_3^2(k')}
    \simeq (SU(5)\,\,\, {\rm symmetric}) 
    + \frac{1}{8 \pi^2}\pmatrix{33/5 \cr 1 \cr -3} \ln\left(\frac{k}{k'}\right).
\label{eq:gc-low}
\end{equation}
This exactly reproduces the MSSM gauge coupling prediction at 
leading-logarithmic level, with the AdS curvature scale $k$ identified as 
the conventional unification scale of $\simeq 10^{16}~{\rm GeV}$ (for more 
details see~\cite{Nomura:2004is}).  There is no proton decay problem 
--- dimension four and five proton decay is suppressed due to the $U(1)_R$ 
symmetry and its unbroken $Z_{2,R}$ subgroup, and dimension six proton 
decay is suppressed because the wavefunctions of the XY gauge fields are 
strongly localized towards the TeV brane and have negligible overlaps with 
matter fields localized on the Planck brane.  Small neutrino masses are also 
naturally obtained through the see-saw mechanism by introducing right-handed 
neutrino superfields $N$ on the Planck brane, together with their Majorana 
masses and Yukawa couplings to the lepton doublets:
\begin{equation}
  S_{\nu} = \int\!d^4x \int_0^{\pi R}\!\!dy \,\, 
    2 \delta(y) \biggl[ \int\!d^2\theta \left( \frac{M_N}{2} N N 
    + y_\nu L N H_u \right) + {\rm h.c.} \biggr].
\label{eq:neutrino}
\end{equation}
Here, $N$ fields carry a $U(1)_R$ charge of $+1$ and are even under the 
$P$ parity.

In the supersymmetric limit, the spectrum of the theory contains exotic 
massless states.  Specifically, the KK spectrum of the gauge tower, $m_n$, 
is approximately given by 
\begin{equation}
  \left\{ \begin{array}{ll} 
    V^{321}: & m_0 = 0, \\
    \{ V^{321}, \Sigma^{321} \}: 
      & m_n \simeq (n-\frac{1}{4})\pi k',
  \end{array} \right.
\qquad
  \left\{ \begin{array}{ll} 
    \Sigma^{\rm XY}: & m_0 = 0, \\
    \{ V^{\rm XY}, \Sigma^{\rm XY} \}: 
      & m_n \simeq (n+\frac{1}{4})\pi k',
  \end{array} \right.
\label{eq:spectrum-gauge}
\end{equation}
where $n = 1,2,\cdots$, so that the zero modes consist of not only the 
321 component of $V$, $V^{321}$, but also the XY component of $\Sigma$, 
$\Sigma^{\rm XY}$, which transforms as $({\bf 3}, {\bf 2})_{-5/6} 
+ ({\bf 3}^*, {\bf 2})_{5/6}$ under 321 (these exotic states, however, 
do not affect the gauge coupling prediction nor lead to rapid proton 
decay as we have seen before).  Once supersymmetry is broken on the 
TeV brane, these exotic states obtain masses~\cite{Nomura:2004is}. 
The fermion component $\lambda'^{\rm XY}$ and the real-scalar 
component $\sigma^{\rm XY}$ in $\Sigma^{\rm XY}$ obtain masses of 
$O(10\!\sim\!100~{\rm TeV})$ through the TeV-brane operators of the 
form $-(\rho e^{-2\pi kR}/4 M_*^2) \int\!d^4\theta\, Z^\dagger Z \,
{\rm Tr}[ {\cal P}[{\cal A}] {\cal P}[{\cal A}]] 
+ \{ (\xi e^{-2\pi kR}/2 M_*) \int\!d^4\theta\, Z^\dagger \,
{\rm Tr}[ {\cal P}[{\cal A}] {\cal P}[{\cal A}]] + {\rm h.c.} \}$, 
where $Z$ represents a chiral superfield responsible for supersymmetry 
breaking, $\langle F_Z \rangle \neq 0$, and ${\cal A}$ is defined by 
${\cal A} \equiv e^{-V}\!(\partial_y e^V) + (\partial_y e^V\!)\,e^{-V}
- \sqrt{2}\, e^V \Sigma\, e^{-V} - \sqrt{2}\, e^{-V} \Sigma^\dagger e^V$. 
(The trace is over the $SU(5)$ space and ${\cal P}[{\cal X}]$ is 
a projection operator: with ${\cal X}$ an adjoint of $SU(5)$, 
${\cal P}[{\cal X}]$ extracts the $({\bf 3}, {\bf 2})_{-5/6} + 
({\bf 3}^*, {\bf 2})_{5/6}$ component of ${\cal X}$ under the 
decomposition to 321.)  The mass of $A_5^{\rm XY}$ is generated 
at one loop through 321 gauge interactions. 

Supersymmetry breaking on the TeV brane also generates masses for 
the 321 gauginos at tree level through the operators
\begin{equation}
  S_{\rm gaugino} = \int\!d^4x \int_0^{\pi R}\!\!dy\,
    2 \delta(y-\pi R) \!\!\sum_{a=1,2,3} \biggl[ 
      -\int\!d^2\theta\, \frac{\zeta_a}{2 M_*} Z \, 
      {\rm Tr}[ {\cal W}_a^\alpha {\cal W}_{a \alpha} ] 
      + {\rm h.c.} \biggr],
\label{eq:gaugino-mass}
\end{equation}
where ${\cal W}_{a \alpha} \equiv -(1/8)\bar{\cal D}^2(e^{-2V} 
{\cal D}_\alpha e^{2V})$ represent field-strength superfields, and 
$a=1,2,3$ denotes $U(1)_Y$, $SU(2)_L$ and $SU(3)_C$, respectively. 
An important point is that the coefficients $\zeta_a$ for the operators 
in Eq.~(\ref{eq:gaugino-mass}) do not respect the $SU(5)$ symmetry, 
as $SU(5)$ is broken to 321 on the TeV brane by boundary conditions. 
These operators, therefore, generate non-universal gaugino masses at 
the TeV scale~\cite{Nomura:2004is}.  This is the 5D realization of 
the condition~(ii) in section~\ref{sec:tuningless}.  The non-universality 
in the gaugino masses is transmitted to the squark and slepton masses, 
which are generated at one loop though the 321 gauge interactions.  This 
allows us to break unwanted unified relations for the scalar masses, 
such as the one in Eq.~(\ref{eq:te-ratio}). 

The Higgs sector of our theory consists essentially of the four Planck-brane 
fields, $S$, $H_u$, $H_d$ and $P'$, and a bulk pedestrian field, $\{ P, 
P^c \}$, which are coupled through the superpotential interactions 
of Eq.~(\ref{eq:Higgs-superpot-5D}).  After supersymmetry is broken, the 
bulk pedestrian field obtains supersymmetry breaking masses through the 
TeV-brane operators
\begin{eqnarray}
  S_P &=& \int\!d^4x \int_0^{\pi R}\!\!dy\,
    2 \delta(y-\pi R)\, e^{-2\pi kR} \int\!d^4\theta\, 
    \biggl[ \frac{1}{2 M_*^2} \Bigl( \zeta_P^{} Z^\dagger P^2
      + \zeta_P^* Z P^{\dagger 2} \Bigr)
    - \frac{\xi_P^{}}{M_*^3} Z^\dagger Z P^\dagger P
\nonumber\\
  && \qquad 
    - \frac{1}{2 M_*^3} \Bigl( \rho_P^{} Z^\dagger Z P^2
      + \rho_P^* Z^\dagger Z P^{\dagger 2} \Bigr)
    - \frac{1}{2 M_*^3} \Bigl( \eta_P^{} Z P^\dagger P
      + \eta_P^* Z^\dagger P^\dagger P \Bigr)  \biggr],
\label{eq:P-mass}
\end{eqnarray}
where $\zeta_P^{}$, $\xi_P^{}$, $\rho_P^{}$ and $\eta_P^{}$ are 
dimensionless parameters.  These operators generate soft supersymmetry 
breaking masses of the form given in Eq.~(\ref{eq:Higgs-soft}), as 
well as the effective supersymmetric mass term for $P$, as given in 
Eq.~(\ref{eq:Pmass-super}).  Imposing $CP$ invariance that is explicitly 
broken only on the Planck brane (e.g. by the Yukawa couplings), we can 
eliminate the supersymmetric $CP$ problem because then we can take the 
basis in which all supersymmetry breaking masses as well as $\lambda$, 
$\eta$ and $\eta'$ are real, which is sufficient to suppress unwanted 
supersymmetric contributions to electric dipole moments.  In the rest 
of the paper we assume the existence of such a basis. 

\subsection{Singlet sector}
\label{subsec:singlet}

The singlet sector of our model has a number of possible variations. 
We have a singlet $S$ that couples to the two Higgs doublets and the 
pedestrian fields $\{P, P^c\}$ and $P'$.  In addition, we can add a set 
of singlet fields, which we collectively call $X$, that transmit the 
scale of the TeV brane to the Planck brane, giving supersymmetric masses 
of order the weak scale to $S$ and $P'$.

The effects of variations of the singlet sector appear essentially only 
in the Higgs sector.  Specifically, the Planck-brane superpotential of 
Eq.~(\ref{eq:Higgs-superpot-5D}) can have additional terms if we extend 
the singlet sector of our theory.  We first note that the superpotential 
of Eq.~(\ref{eq:Higgs-superpot-5D}) can have an additional term 
\begin{equation}
  \delta S_{\rm Higgs} = \int\!d^4x \int_0^{\pi R}\!\!dy \,\, 
    2 \delta(y) \biggl[ \int\!d^2\theta\, \frac{\kappa}{3} S^3 
    + {\rm h.c.} \biggr].
\label{eq:Higgs-superpot-5D-add-1}
\end{equation}
This term explicitly breaks the $U(1)_R$ symmetry to the $Z_{4,R}$ subgroup, 
but this $Z_{4,R}$ is still sufficient to forbid  dangerous operators 
such as the ones leading to a large Higgs mass and dimension four and five 
proton decay.  The coupling $\kappa$ cannot be very large so that it does 
not give too large of an asymptotically non-free contribution to the evolution 
of $\lambda$.  This constrains the size of $\kappa$ as $\kappa \simlt 0.3$ 
($0.4$) for $\lambda \simgt 0.7$ ($0.6$).  While we mostly concentrate 
on the case with $\kappa = 0$, the case with $\kappa \neq 0$ will also 
be considered when we discuss electroweak symmetry breaking later.

The Higgs-sector superpotential can also contain a supersymmetric mass 
term for $S$:
\begin{equation}
  \delta S_{\rm Higgs} = \int\!d^4x \int_0^{\pi R}\!\!dy \,\, 
    2 \delta(y) \biggl[ \int\!d^2\theta\, \frac{M_S}{2} S^2 
    + {\rm h.c.} \biggr],
\label{eq:Higgs-superpot-5D-add-2}
\end{equation}
where $M_S$ is a mass parameter of order the weak scale.  This term can 
naturally arise if there is a singlet field $\{X, X^c\}$ in the bulk 
that couples both to the TeV brane and the $S$ field on the Planck brane. 
Suppose that $\{X, X^c\}$ has a bulk mass $c_X \simeq 1/2$ and couples 
to $S$ on the Planck brane as
\begin{equation}
  \delta S = \int\!d^4x \int_0^{\pi R}\!\!dy \,\, 
    2 \delta(y) \biggl[ \int\!d^2\theta\, \frac{\lambda_S}{2} X S^2 
    + {\rm h.c.} \biggr].
\label{eq:Higgs-superpot-5D-add-3}
\end{equation}
Then, if the TeV-brane physics gives the VEV of the $X$ field, say through 
the superpotential term $\delta(y-\pi R) \int\!d^2\theta\, Y (X^2 - M^2)$, 
where $M$ is some mass parameter and $Y$ is a Lagrange multiplier, the 
generated mass for $S$ on the Planck brane, $M_S = \lambda_S \langle 
X \rangle$ is naturally of order the weak scale (this has been used 
in~\cite{Goldberger:2002pc} to generate a weak-scale $\mu$ term for 
the Higgs doublets on the Planck brane).  Here we assume that the VEV 
for $F_X$, which can be generated through supersymmetry breaking effects, 
is parametrically suppressed.  Similarly, the $\{X, X^c\}$ field could 
also generate a supersymmetric mass term for $P'$ on the Planck brane:
\begin{equation}
  \delta S_{\rm Higgs} = \int\!d^4x \int_0^{\pi R}\!\!dy \,\, 
    2 \delta(y) \biggl[ \int\!d^2\theta\, \frac{M_{P'}}{2} P'^2 
    + {\rm h.c.} \biggr],
\label{eq:Higgs-superpot-5D-add-4}
\end{equation}
where $M_{P'}$ is naturally of order the weak scale.%
\footnote{To preserve $Z_{4,R}$ on the Planck brane (up to the 
weak-scale effects), $Z_{4,R}$ charges of $\{X, X^c\}$'s generating 
Eqs.~(\ref{eq:Higgs-superpot-5D-add-2}) and (\ref{eq:Higgs-superpot-5D-add-4}) 
need to be different.}

The terms in Eqs.~(\ref{eq:Higgs-superpot-5D-add-1},%
~\ref{eq:Higgs-superpot-5D-add-2}) affect the phenomenology of the Higgs 
sector, including electroweak symmetry breaking and the chargino/neutralino 
spectrum.  The presence of the term in Eq.~(\ref{eq:Higgs-superpot-5D-add-4}) 
could be important for keeping $P$ parity unbroken.  In the analysis 
in later sections, we will treat $\kappa$, $M_S$ and $M_{P'}$ as free 
parameters with $M_S$ and $M_{P'}$ of order the weak scale.  It should 
be remembered, however, that these terms can naturally arise without 
introducing any hierarchically small parameters.

\subsection{Supersymmetry-breaking parameters}
\label{subsec:formula}

Here we present simple, approximate formulae for the gaugino and scalar 
masses, using the holographic 4D description of our theory.  These 
formulae were derived in~\cite{Nomura:2004zs} and will be used in the 
analysis in later sections. 

Let us suppose that supersymmetry breaking on the TeV brane is not very 
strong (i.e. the parameters $\zeta_a F_Z$ are not large compared with 
the appropriately rescaled curvature scale), which is the case we 
concentrate on in this paper.  In this case, the gaugino and scalar 
masses are generated in the 4D picture by 321 gauge interactions that 
link them to the DSB sector.  Using a scaling argument based on the 
large-$N$ expansion~\cite{'tHooft:1973jz}, the masses for the gauginos, 
$M_a \equiv m_{\lambda^{321}_a}$ ($a=1,2,3$), are estimated as
\begin{equation}
  M_a \simeq g_a^2 \frac{\tilde{N}}{16\pi^2}\, \hat{\zeta}_a\, m_\rho,
\label{eq:gaugino-masses-1}
\end{equation}
where $g_a$ are the 4D 321 gauge couplings, $\hat{\zeta}_a$ are 
dimensionless parameters of $O(1)$ that depend on the gauge group, 
$\tilde{N}$ is the size of the DSB gauge group $G$, and $m_\rho$ 
is the typical mass scale of resonances in the DSB sector 
(i.e. the bound states arising from the IR dynamics of $G$).%
\footnote{In a theory where $G$ is almost conformal above the dynamical 
scale $\Lambda$, the parameter $\tilde{N}$ may actually represent the 
{\it square} of the number of ``colors'' of $G$, and not the number of 
``colors'' itself.  Discussions on this and related issues in the AdS/CFT 
correspondence can be found, for example, in Ref.~\cite{Burdman:2003ya}.}
Similarly, the squared masses for the scalars, $m_{\tilde{f}}^2$, are 
estimated as
\begin{equation}
  m_{\tilde{f}}^2 \simeq \sum_{a=1,2,3} 
    \frac{g_a^4 C_a^{\tilde{f}}}{16\pi^2} 
    \frac{\tilde{N}}{16\pi^2}\, \hat{\zeta}_a^2\, m_\rho^2,
\label{eq:scalar-masses-1}
\end{equation}
where $\tilde{f} = \tilde{q}, \tilde{u}, \tilde{d}, \tilde{l}, \tilde{e}$ 
represents the MSSM squarks and sleptons, and $C_a^{\tilde{f}}$ are 
the group theoretical factors given by $(C_1^{\tilde{f}}, C_2^{\tilde{f}}, 
C_3^{\tilde{f}}) = (1/60,3/4,4/3)$, $(4/15,0,4/3)$, $(1/15,0,4/3)$, 
$(3/20,3/4,0)$ and $(3/5,0,0)$ for $\tilde{f} = \tilde{q}, \tilde{u}, 
\tilde{d}, \tilde{l}$ and $\tilde{e}$, respectively.  Since these masses 
are generated through gauge interactions, they are flavor universal and 
the supersymmetric flavor problem is absent.  Note that, because of the 
presence of $\hat{\zeta}_a^2$ in Eq.~(\ref{eq:scalar-masses-1}), which 
is required to correctly pick up the effect of supersymmetry breaking, 
squark and slepton masses do not obey relations arising from the $SU(5)$ 
symmetry.

The gaugino and scalar masses in Eqs.~(\ref{eq:gaugino-masses-1},~%
\ref{eq:scalar-masses-1}) can be expressed in terms of 5D quantities 
in the following way.  Let us first identify the relevant parameters in 5D. 
In our 5D theory, the tree-level brane-localized gauge kinetic terms are 
assumed to be small (this assumption does not necessarily have to be made 
for the Planck-brane localized terms in the case that $SU(5)$ is broken 
by a Higgs VEV on the Planck brane). Now, the relevant parameters for 
the superparticle masses are the coefficients of the bulk and brane gauge 
kinetic terms renormalized at the scale $k'$, measured in terms of the 
4D metric $\eta_{\mu\nu}$.  This implies that, while the coefficients 
for the TeV-brane gauge kinetic terms can still be regarded as small, 
the (renormalized) coefficients for the Planck-brane gauge kinetic terms 
are not, because they are enhanced by a large logarithm, $\ln(k/k')$, 
through their RG evolution from the scale $k$ down to the scale $k'$. 
We can thus write the gauge kinetic part of the Lagrangian, renormalized 
at the 4D scale $k'$, as
\begin{equation}
  {\cal L}_{\rm ren.\, 5D} 
    \approx -\frac{1}{4g_B^2} F_{\mu\nu} F^{\mu\nu} 
    - 2 \delta(y) \frac{1}{4\tilde{g}_{0,a}^2} {F^a}_{\mu\nu} {F^a}^{\mu\nu},
\label{eq:gen-kin}
\end{equation}
where $g_B$ is the $SU(5)$-invariant 5D gauge coupling and 
$1/\tilde{g}_{0,a}^2$ the renormalized coefficients for the Planck-brane 
gauge kinetic terms.  The 4D gauge couplings $g_a$ are then given by 
\begin{equation}
  \frac{1}{g_a^2} = \frac{\pi R}{g_B^2} + \frac{1}{\tilde{g}_{0,a}^2},
\label{eq:4D-gc}
\end{equation}
at the scale $k'$. Identifying the contribution to $1/g_a^2$ from the bulk, 
$\pi R/g_B^2$, as the RG contribution to the 321 gauge couplings from the 
DSB sector, $(\tilde{N}/16\pi^2)\ln(k/k')$, we obtain the correspondence 
relation
\begin{equation}
  \frac{\tilde{N}}{16\pi^2} \approx \frac{1}{g_B^2 k}.
\label{eq:corresp-1}
\end{equation}
The scale for the resonance masses, $m_\rho$, corresponds in the 5D picture 
to the scale for the KK masses, $\pi k'$, so 
\begin{equation}
  m_\rho \approx \pi k'.
\label{eq:corresp-2}
\end{equation}
The parameter $\hat{\zeta}_a$ can then be read off by matching the gaugino 
mass expression of Eq.~(\ref{eq:gaugino-masses-1}) to the approximate 
tree-level gaugino mass expression in 5D, $g_a^2 (\zeta_a F_Z/M_*) (k'/k)$, 
as 
\begin{equation}
  \hat{\zeta}_a \approx \frac{\zeta_a g_B^2 F_Z}{\pi M_*},
\label{eq:corresp-3}
\end{equation}
where the parameters $\zeta_a$ and $M_*$ appear in Eq.~(\ref{eq:gaugino-mass}) 
and $F_Z$ is the VEV of the highest component of the chiral superfield $Z$.%
\footnote{The definition of $F_Z$ in this paper is given as follows.  In the 
normalization where the kinetic term of $Z$ is canonically normalized in 4D, 
$F_Z$ is defined by $F_Z = e^{\pi kR} \partial Z/\partial \theta^2|_{\theta 
= \bar{\theta} = 0}$.  The natural size for $F_Z$ is then of order $k^2 
\sim M_*^2 \sim M_{\rm Pl}^2$ (no exponential suppression factor).}
If we assume that the sizes of various parameters are given by naive 
dimensional analysis~\cite{Manohar:1983md}, we obtain $g_B^2 \simeq 
16\pi^3/M_*$, $F_Z \simeq M_*^2/4\pi$ and $\zeta_a \simeq 1/4\pi$, and 
we find that $\hat{\zeta}_a$ are in fact of $O(1)$. 

Using the correspondence relations Eqs.~(\ref{eq:corresp-1},~%
\ref{eq:corresp-2},~\ref{eq:corresp-3}), we finally obtain the 
following simple formulae for the gaugino and scalar masses:
\begin{equation}
  M_a = g_a^2\, \frac{\zeta_a F_Z}{M_*} \frac{k'}{k},
\label{eq:gaugino-masses-2}
\end{equation}
and 
\begin{equation}
  m_{\tilde{f}}^2 
  = \gamma\!\! \sum_{a=1,2,3} \frac{g_a^4 C_a^{\tilde{f}}}{16 \pi^2}\, 
    (g_B^2 k) \Biggl( \frac{\zeta_a F_Z}{M_*} \frac{k'}{k} \Biggr)^2,
\label{eq:scalar-masses-2}
\end{equation}
where $g_a$ are the 4D gauge couplings given by Eq.~(\ref{eq:4D-gc}) and 
$\gamma$ is a numerical coefficient of $O(1)$.  The quantity $M_{{\rm SUSY},a} 
\equiv (\zeta_a F_Z/M_*)(k'/k)$, which sets the overall mass scale 
in Eqs.~(\ref{eq:gaugino-masses-2},~\ref{eq:scalar-masses-2}), is of 
$O(M_* e^{-\pi kR}/16\pi^2)$, and so is naturally of $O(100~{\rm 
GeV}\!\sim\!1~{\rm TeV})$ for $k' \simeq (10\!\sim\!100)~{\rm TeV}$. 
These expressions can be checked (numerically) by 5D calculations, 
as was done in Ref.~\cite{Nomura:2003qb} (for $\zeta_1 = \zeta_2 
= \zeta_3$, Eqs.~(\ref{eq:gaugino-masses-2},~\ref{eq:scalar-masses-2}) 
reproduce the mass spectrum given in~\cite{Nomura:2003qb}).  The numerical 
coefficient $\gamma$ takes values $\gamma \simeq (5\!\sim\!6)$, 
and is not very sensitive to the parameters of the model.

\section{Electroweak Symmetry Breaking}
\label{sec:ewsb}

In this section we study electroweak symmetry breaking in our theory. 
We show that the correct value for the electroweak scale is obtained 
without severe fine-tuning.  We also work out the superparticle spectrum 
of the theory and discuss its generic features.  Some phenomenological 
analyses, especially those for the neutralino and pedestrian sectors, 
will be deferred to the next section.

\subsection{Parameters in the Higgs sector}
\label{subsec:para-Higgs}

The Higgs sector of our theory consists of the Planck-brane fields $S$, 
$H_u$, $H_d$ and $P'$, and the bulk pedestrian field $\{ P, P^c \}$, together 
with the interactions of Eqs.~(\ref{eq:Higgs-superpot-5D},~\ref{eq:P-mass}). 
There can also be additional terms Eqs.~(\ref{eq:Higgs-superpot-5D-add-1},%
~\ref{eq:Higgs-superpot-5D-add-2},~\ref{eq:Higgs-superpot-5D-add-4}). 
After dimensional reduction, the Higgs sector consists of $S$, $H_u$, 
$H_d$, $P'$ and the zero mode of $P$, which have the interactions of 
the form of Eqs.~(\ref{eq:Higgs-superpot},~\ref{eq:Higgs-Z}) but with 
Eq.~(\ref{eq:Higgs-superpot}) having the additional piece $\delta W_H 
= \eta' S P P' + (h/2) S P'^2$ (and the pieces coming from 
Eqs.~(\ref{eq:Higgs-superpot-5D-add-1},~\ref{eq:Higgs-superpot-5D-add-2},%
~\ref{eq:Higgs-superpot-5D-add-4})).  After integrating out the $P$ field, 
which is expected to be somewhat heavier because of the supersymmetric 
mass term at tree level, we obtain the effective Higgs sector, given 
by Eqs.~(\ref{eq:Higgs2-W},~\ref{eq:Higgs2-L}) but supplemented by 
additional terms involving the $P'$ fields:
\begin{equation}
  W = \lambda S H_u H_d + L_S^2 S + \frac{h}{2} S P'^2 
      + \frac{M_{P'}}{2} P'^2,
\label{eq:Higgs_eff-W}
\end{equation}
\begin{eqnarray}
  {\cal L}_{\rm soft} 
    &=& - m_{H_u}^2 |H_u|^2 - m_{H_d}^2 |H_d|^2 
        - m_S^2 |S|^2 - m_{P'}^2 |P'|^2 
\nonumber\\
    &&  - \Bigl( \lambda A_\lambda S H_u H_d + \frac{h}{2} A_\lambda S P'^2 
        + L_S^2 C_S S + {\rm h.c.} \Bigr).
\label{eq:Higgs_eff-L}
\end{eqnarray}
Here, we have included a term coming from 
Eq.~(\ref{eq:Higgs-superpot-5D-add-4}), which could potentially 
be present. The Higgs potential we study is then given, for 
$\langle P' \rangle = 0$, by Eq.~(\ref{eq:Higgs-Pot}).  In the case 
that the superpotential terms in Eqs.~(\ref{eq:Higgs-superpot-5D-add-1},%
~\ref{eq:Higgs-superpot-5D-add-2}) are added, $W$ and ${\cal L}_{\rm soft}$ 
in Eqs.~(\ref{eq:Higgs_eff-W},~\ref{eq:Higgs_eff-L}) are supplemented by 
the terms 
\begin{eqnarray}
  \delta W &=& \frac{M_S}{2} S^2 + \frac{\kappa}{3} S^3,
\label{eq:Higgs_eff-W-add}\\
  \delta {\cal L}_{\rm soft} 
    &=& - \Bigl( M_S A_\lambda S^2 + \kappa A_\lambda S^3 
        + {\rm h.c.} \Bigr),
\label{eq:Higgs_eff-L-add}
\end{eqnarray}
where the terms in $\delta {\cal L}_{\rm soft}$ arise after integrating 
out the $F_S$ field, through the term $(F_S^\dagger S + {\rm h.c.})$ 
radiatively generated via the diagram in Fig.~\ref{fig:loop-Spot}c.

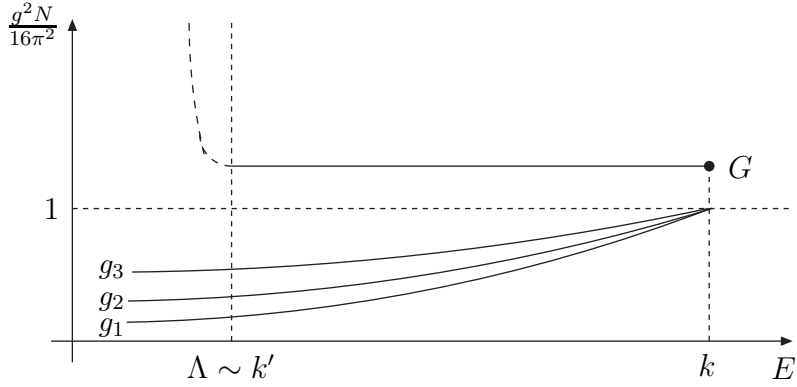
\begin{figure}[t]
\begin{center}
\begin{picture}(300,143)(-15,-20)
  \DashCArc(254,120)(220,180,193){3} 
  \DashCArc(50,78)(12,180,270){3} 
  \Line(50,66)(230,66)  \Text(238,66)[l]{$G$}
  \CArc(-2,657)(650,271.1,291)  \Text(10,5)[r]{\small $g_1$}
  \CArc(-3,815)(800,271,287)    \Text(10,15)[r]{\small $g_2$}
  \CArc(1,1126)(1100,270.6,282) \Text(10,27)[r]{\small $g_3$}
  \DashLine(-10,50)(260,50){2} \Text(-15,50)[r]{$1$}
  \DashLine(50,0)(50,120){2} \Text(50,-5)[t]{$\Lambda \sim k'$}
  \DashLine(230,66)(230,0){2} \Vertex(230,66){2} \Text(230,-5)[t]{$k$}
  \LongArrow(-10,-8)(-10,120) \Text(-15,120)[r]{$\frac{g^2 N}{16\pi^2}$}
  \LongArrow(-18,0)(260,0) \Text(259,-6)[t]{$E$}
\end{picture}
\caption{Schematic description for the evolution of the gauge couplings 
 in our theory.} 
\label{fig:couplings}
\end{center}
\end{figure}

To discuss electroweak symmetry breaking quantitatively, we need the 
sizes of parameters appearing in the Higgs potential.  In particular, 
we need to know the sizes of the coupling $\lambda$ (and $\kappa$) and 
the dimensionful parameters $L_S^2$, $m_{H_u}^2$, $m_{H_d}^2$, $m_S^2$, 
$A_\lambda$ and $L_S^2 C_S$ (and $M_S$).  We first consider the coupling 
$\lambda$.  The upper bound on the size of $\lambda$ at the weak scale 
is given by the condition that it does not blow up below the unification 
scale of $\simeq 10^{16}~{\rm GeV}$.  To derive the bound, therefore, 
we have to evolve parameters from a high scale down to the weak scale. 
For this purpose it is useful to consider the theory in the holographic 
4D picture.  The 4D theory is defined at the UV cutoff scale of order 
$k \sim M_{\rm pl}$ and contains a sector (DSB sector) that has a gauge 
interaction with the group $G$, whose coupling $\tilde{g}$ evolves very 
slowly over a wide energy interval below $k$.  Denoting the size of the 
group $G$ to be $\tilde{N}$, the correspondence is given by $\tilde{g}^2 
\tilde{N}/16\pi^2 \approx M_*/\pi k$ and $\tilde{N} \approx 16\pi^2/g_B^2 
k$ (so $\tilde{g} \simeq 4\pi$ and $\tilde{N} \simgt 1$ here, see also 
Eq.~(\ref{eq:corresp-1})).  The bulk gauge symmetry and the Planck-brane 
boundary conditions in the 5D theory imply that the $G$ gauge sector 
possesses a global $SU(5)$ symmetry whose $SU(3)_C \times SU(2)_L \times 
U(1)_Y$ subgroup is explicitly gauged.  Since the DSB sector is charged 
under 321, it contributes to the running of the 321 gauge couplings, 
$g_a$.  The RG equations for the 321 gauge couplings in the 4D picture 
are thus given by 
\begin{equation}
  \frac{d}{d\ln\mu} \left( \frac{1}{g_a^2} \right) 
    = -\frac{1}{8 \pi^2} \left( b^{\rm MSSM}_a + b^{\rm DSB} \right),
\label{eq:RG-gauge}
\end{equation}
where $(b^{\rm MSSM}_1, b^{\rm MSSM}_2, b^{\rm MSSM}_3)=(33/5, 1, -3)$ 
are the MSSM beta-function coefficients, and $b^{\rm DSB}$ represents 
the contribution from the DSB sector, which is given by $b^{\rm DSB} = 
8\pi^2/g_B^2 k$ in terms of the 5D quantities.  Because of the global 
$SU(5)$ symmetry of the DSB sector, $b^{\rm DSB}$ is universal, i.e. 
$b^{\rm DSB}$ does not depend on $a$.  In the case that tree-level 
Planck-brane kinetic terms are small, as must be the case for the 
boundary-condition $SU(5)$ breaking on the Planck brane, the 321 gauge 
couplings in the 4D picture, $g_a$, approach a Landau pole at the 
scale $\sim k$, and the DSB contribution is determined as $b^{\rm CFT} 
\simeq 4.8$.  For the Higgs $SU(5)$-breaking case, $g_a$ at the scale 
$k$ can be smaller, so that $b^{\rm CFT} \simlt 4.8$.  A schematic 
description for the evolution of the gauge couplings in the 4D picture 
is given in Fig.~\ref{fig:couplings}.

The modes localized to the Planck brane correspond to the fields singlet 
under $G$ (elementary fields).  In particular, the MSSM quarks, leptons and 
the $S$, $H_u$, $H_d$ and $P'$ fields are elementary states.  The 4D theory 
also contains the $P$ field as an elementary field, which interacts with 
the DSB sector through the interaction of the form
\begin{equation}
  W = \lambda_P P \cdot {\cal O}_P,
\label{eq:P-interaction}
\end{equation}
where ${\cal O}_P$ is a $G$-singlet operator of the DSB sector, whose 
dimension is close to 2: $[{\cal O}_P] \simeq 2$, and $\lambda_P$ is a 
coupling which is almost marginal.  The KK states for the $\{P, P^c\}$ 
field in the 5D picture correspond to mixtures of the elementary $P$ state 
and the composite states interpolating the operator ${\cal O}_P$.  The 
interactions among the elementary fields in the Higgs sector are given by
\begin{equation}
  W = \lambda S H_u H_d + \frac{\eta}{2} S P^2 
      + \eta' S P P' + \frac{h}{2} S P'^2,
\label{eq:Higgs-superpot-4D}
\end{equation}
where we have set $M_S = \kappa = 0$, for simplicity.  According to 
the supersymmetric non-renormalization theorem, the couplings $\lambda$, 
$\eta$, $\eta'$ and $h$ run only through anomalous dimensions of the 
$S$, $H_u$, $H_d$, $P$ and $P'$ fields, which receive contributions 
from the gauge and Yukawa interactions as well as the interactions in 
Eqs.~(\ref{eq:P-interaction},~\ref{eq:Higgs-superpot-4D}).  Let us now 
write down the RG equations for $\eta$ and $\eta'$, incorporating the 
contribution from the DSB sector, Eq.~(\ref{eq:P-interaction}), along the 
lines discussed in~\cite{Birkedal:2004zx}.  Denoting the 5D couplings of 
the superpotential terms $S P^2$ and $S P P'$ ($\eta$ and $\eta'$ appearing 
in Eq.~(\ref{eq:Higgs-superpot-5D})) as $\hat{\eta}$ and $\hat{\eta}'$, 
respectively, the couplings $\eta$ and $\eta'$ in the 4D picture ($\eta$ 
and $\eta'$ appearing in Eq.~(\ref{eq:Higgs-superpot-4D})) at the RG scale 
$\mu$ are given by
\begin{equation}
  \eta(\mu) = \frac{\hat{\eta} M}{\sqrt{Z_S(\mu)\, Z_P(\mu)^2}}, \qquad
  \eta'(\mu) = \frac{\hat{\eta}' \sqrt{M}}
               {\sqrt{Z_S(\mu)\, Z_P(\mu)\, Z_{P'}(\mu)}},
\label{eq:RG-eta}
\end{equation}
where $Z_S(\mu)$, $Z_P(\mu)$ and $Z_{P'}(\mu)$ are wavefunction 
renormalization factors for the $S$, $P$ and $P'$ fields, which obey 
the RG equations
\begin{eqnarray}
  \frac{d\ln Z_S}{d\ln\mu} &=& -\frac{1}{8\pi^2} \biggl( 
    2 \lambda^2 + \frac{1}{2}\frac{\hat{\eta}^2 M^2}{Z_S Z_P^2} 
    + \frac{\hat{\eta}'^2 M}{Z_S Z_P Z_{P'}} + \frac{1}{2} h^2 \biggr),
\label{eq:RG-ZS} \\
  \frac{d\ln Z_P}{d\ln\mu} &=& -\frac{1}{8\pi^2} 
    \biggl( \frac{\hat{\eta}^2 M^2}{Z_S Z_P^2} 
    + \frac{\hat{\eta}'^2 M}{Z_S Z_P Z_{P'}} \biggr) - \frac{M}{k Z_P},
\label{eq:RG-ZP} \\
  \frac{d\ln Z_{P'}}{d\ln\mu} &=& -\frac{1}{8\pi^2} \biggl( 
    \frac{\hat{\eta}'^2 M}{Z_S Z_P Z_{P'}} + h^2 \biggr),
\label{eq:RG-ZPp}
\end{eqnarray}
(the 5D couplings $\hat{\eta}$ and $\hat{\eta}'$ do not run: 
$d\hat{\eta}/d\ln\mu = d\hat{\eta}'/d\ln\mu = 0$).  Here, $M$ is 
a spurious parameter relating the 5D and 4D $P$ fields: $P_{\rm 5D} 
= \sqrt{M} P_{\rm 4D}$;  the physics should not depend on it. 
The boundary conditions for $Z_S$, $Z_P$ and $Z_{P'}$ are given by
\begin{equation}
  Z_S(k)=Z_{0,S},\qquad Z_P(k)=M Z_{0,P},\qquad Z_{P'}(k)=Z_{0,P'},
\end{equation}
where $Z_{0,S}$, $Z_{0,P}$ and $Z_{0,P'}$ represent tree-level kinetic 
terms, $2\delta(y) \int\!d^4\theta (Z_{0,S} S^\dagger S + Z_{0,P} 
P^\dagger P + Z_{0,P'} P'^\dagger P')$, localized on the Planck brane. 
For the case of strong coupling at the fundamental scale, the UV parameters 
are estimated as $\hat{\eta} \approx 4\pi^2/M_*$, $\hat{\eta}' \approx 
4\pi\sqrt{\pi/M_*}$, $Z_{0,S} \approx Z_{0,P'} \approx 1$ and $Z_{0,P} 
\approx \pi/M_*$ (using naive dimensional analysis), so that $\eta(k) 
\approx \eta'(k) \approx 4\pi$.  We can also easily see that $Z_S(\mu), 
Z_{P'}(\mu) \propto M^0$ and $Z_P(\mu) \propto M$ at an arbitrary scale 
$\mu$, so that $\eta(\mu)$ and $\eta'(\mu)$ in fact do not depend on 
the spurious parameter $M$. 

Solving the RG equations for $\eta$ and $\eta'$ given by 
Eqs.~(\ref{eq:RG-eta}~--~\ref{eq:RG-ZPp}), we find that the values of 
$\eta$ and $\eta'$ are suppressed at low energies due to the contribution 
from the DSB sector, represented as the second term in the right-hand-side 
of Eq.~(\ref{eq:RG-ZP}).  This is simply the 4D realization of the fact 
that the couplings $\eta$ and $\eta'$ receive volume suppressions in the 5D 
theory because the $P$ field propagates in the bulk.  This is advantageous 
for the evolution of $\lambda$, since the RG equation for $\lambda$ is 
given by
\begin{eqnarray}
  \frac{d\, \lambda}{d\ln\mu} &=& \frac{\lambda}{16\pi^2} \biggl( 
    4 \lambda^2 + \frac{1}{2} \eta^2 + \eta'^2 + \frac{1}{2} h^2 
    + 3 y_t^2 - \frac{3}{5} g_1^2 - 3 g_2^2 \biggr),
\label{eq:RG-lambda}
\end{eqnarray}
so that smaller $\eta$ and $\eta'$ help to obtain larger values for 
$\lambda$ at low energies. Here, $y_t$ is the top Yukawa coupling, which 
obeys the RG equation
\begin{equation}
  \frac{d\, y_t}{d\ln\mu} = \frac{y_t}{16\pi^2} \biggl( 
    6 y_t^2 + \lambda^2 - \frac{13}{15} g_1^2 - 3 g_2^2 
    - \frac{16}{3} g_3^2 \biggr),
\label{eq:RG-yt}
\end{equation}
and we have neglected the bottom and tau Yukawa couplings (since we are 
interested in a small $\tan\beta$ region) as well as the Yukawa couplings 
for the first two generations. 

Our theory has the following three features that allow larger values 
of $\lambda$ at the weak scale compared with the conventional NMSSM. 
(i) The theory has larger gauge couplings at high energies than 
the conventional MSSM/NMSSM (see Eq.~(\ref{eq:RG-gauge})) so that 
$\lambda$ is less asymptotically non-free (due to the last two terms in 
Eq.~(\ref{eq:RG-lambda})).  (ii) Larger gauge couplings at high energies 
give smaller values for the top Yukawa coupling at high energies with 
a fixed value of the top quark mass (due especially to the last term in 
Eq.~(\ref{eq:RG-yt})), which reduces the asymptotically non-free contribution 
to the $\lambda$ running from the top Yukawa coupling given by the third 
term of Eq.~(\ref{eq:RG-lambda}).  (iii) The superpotential Higgs cubic 
coupling $\kappa$ does not have to be large ($\kappa$ can even be zero), 
as the stabilization of the VEV of $S$ does not require this term in 
our theory.  This eliminates a potentially large asymptotically non-free 
contribution coming from non-zero $\kappa$, which would add a term $\lambda 
\kappa^2/8\pi^2$ to the right-hand-side of Eq.~(\ref{eq:RG-lambda}). The 
first two features (i) and (ii) were considered earlier in~\cite{Kane:1992kq}. 
Note that, unlike $\kappa$ in the NMSSM, the coupling $h$ need not be 
sizable so that its effect on the evolution of $\lambda$ can be quite 
small.  The point~(ii) is especially significant, which gives values of 
the top Yukawa coupling at the scale $k$ as small as $y_t(k) = O(10^{-2})$. 
Together with the fact that the couplings $\eta$ and $\eta'$ are strongly 
asymptotically non-free, we find that we can easily obtain values of $\lambda$ 
as large as 
\begin{equation}
  \lambda \simeq 0.8
\label{eq:lambda-value}
\end{equation}
at the weak scale, and we will mainly use this value in our analysis 
below.  The couplings $\eta$ and $\eta'$ at the weak scale can be as 
large as $\eta \simeq 0.1$ and $\eta' \simeq 0.3$, without affecting the 
result for the maximum value of $\lambda$ in Eq.~(\ref{eq:lambda-value}). 
A value for the coupling $h$ smaller than about $\simeq 0.4$ does 
not affect the value in Eq.~(\ref{eq:lambda-value}), either. 
For $\kappa \neq 0$, RG equations are modified such that $h^2$ in 
Eqs.~(\ref{eq:RG-ZS},~\ref{eq:RG-lambda}) is replaced by $h^2+4 
\kappa^2$, and the RG equation for $\kappa$ is given by $d\kappa/d\ln\mu 
= (3\kappa/16\pi^2)(2\kappa^2-8\pi^2 (d\ln Z_S/d\ln\mu))$.  The condition 
for Eq.~(\ref{eq:lambda-value}) to hold is then given by $h^2 + 4\kappa^2 
\simlt (0.4)^2$ at the weak scale.

The value of $y_t$ should not be very large so as not to give a large 
asymptotically non-free contribution to the evolution of $\lambda$.  This 
gives a lower bound on $\tan\beta$, roughly given by $\tan\beta \simgt 1.7$ 
for $\lambda \simeq 0.8$ ($\tan\beta \simgt 1.4$ for $\lambda \simeq 0.7)$. 
Note that this bound does not come from the requirement of the perturbativity 
of $y_t$ up to the unification scale $\approx k$, but instead comes from 
the requirement of having large enough $\lambda$ in the IR and thus a large 
enough physical Higgs-boson mass.  In fact, for these values of $\tan\beta$, 
the top Yukawa coupling is strongly asymptotically free and thus perturbative 
up to the scale $k$, due to large contributions from the gauge couplings to 
the evolution of $y_t$. 

Next we consider the sizes of the dimensionful parameters $L_S^2$, 
$L_S^2 C_S$, $A_\lambda$, $m_S^2$ and $m_{P'}^2$ appearing in 
Eqs.~(\ref{eq:Higgs_eff-W},~\ref{eq:Higgs_eff-L}).  These parameters 
are generated through the diagrams of Fig.~\ref{fig:loop-Spot} and 
similar diagrams, from supersymmetry breaking masses for $P$'s given 
by Eq.~(\ref{eq:P-mass}).  Their values are given by the expressions like 
Eqs.~(\ref{eq:LS2}~--~\ref{eq:mS2}), but now with the cutoff $\Lambda$ 
identified as the KK mass scale, $\Lambda = O(k')$, due to locality 
in the 5D theory.  For $\eta' \simgt \eta$, they are given by
\begin{eqnarray}
  L_S^2 &\approx& 
    -\frac{\eta}{16\pi^2} M_P^* B_P^* \ln\left(\frac{k'}{|M_P|}\right),
\label{eq:LS2-5D} \\
  L_S^2 C_S &\approx& 
    -\frac{\eta}{16\pi^2} A_\eta M_P^* B_P^* 
        \ln\left(\frac{k'}{|M_P|}\right),
\label{eq:LS2CS-5D} \\
  A_\lambda &\approx& 
    -\frac{|\eta'|^2}{8\pi^2} A_\eta \ln\left(\frac{k'}{|M_P|}\right),
\label{eq:Alambda-5D} \\
  m_S^2 \:\:\:\approx\:\:\:  m_{P'}^2 &\approx& 
    -\frac{|\eta'|^2}{8\pi^2} m_P^2 \ln\left(\frac{k'}{|M_P|}\right).
\label{eq:mS2-mPp2-5D}
\end{eqnarray}
We will take the values of $L_S^2$, $L_S^2 C_S$, $A_\lambda$ and $m_S^2$ 
essentially as free parameters in our analysis of the Higgs potential, 
for the following reasons.  First of all, the four parameters $L_S^2$, 
$L_S^2 C_S$, $A_\lambda$ and $m_S^2$ depend on the quantities $\eta$, 
$\eta'$, $M_P B_P$, $A_\eta$ and $m_P^2$, which are free parameters 
of the theory.  Therefore, in general there is no particular relation 
among the parameters $L_S^2$, $L_S^2 C_S$, $A_\lambda$ and $m_S^2$. 
One may still worry that there may be upper bounds on the sizes of these 
parameters for a given value of $k'$, especially because low-energy values 
of $\eta$ and $\eta'$ are bounded as $\eta \simlt 0.1$ and $\eta' \simlt 
0.3$.  In fact, for the lowest KK gauge-boson mass of order $200~{\rm TeV}$, 
which allows $|m_P^2|$ as large as $\approx (20~{\rm TeV})^2$, the value 
of $m_S^2$ is bounded as $m_S^2 \simlt (800~{\rm GeV})^2$.  Similar bounds 
also apply to the other parameters, giving $L_S^2 \simlt (600~{\rm GeV})^2$, 
$L_S^2 C_S \simlt (2~{\rm TeV})^3$ and $A_\lambda \simlt 20~{\rm GeV}$ 
for the same value of $k'$ (or the KK mass).  Since we are interested in the
region where the lowest KK gauge-boson mass is smaller than a few hundred 
TeV, i.e. $k' \simlt 100~{\rm TeV}$, we expect that these parameters should 
not take values far in excess of  the above bounds.  In fact, we will 
see in the next subsection that correct electroweak symmetry breaking 
is obtained for the values of $A_\lambda$ and $m_S^2$ almost saturating 
these bounds (see Table~\ref{table:points}).  The values of $L_S^2$ and 
$L_S^2 C_S$ must be somewhat suppressed compared with the bounds, but 
this can easily be attained by taking the coupling $\eta$ small, $\eta 
\approx O(10^{-3}\!\sim\!10^{-2})$ (see Eqs.~(\ref{eq:LS2-5D}--%
\ref{eq:mS2-mPp2-5D})).  Note that small values of $\eta$ are natural 
because $\eta$ is a superpotential coupling located on the UV brane.%
\footnote{It is non-trivial, in fact, that we obtain correct electroweak 
symmetry breaking in the parameter region in which only $\eta$ is small 
and all the unprotected IR-brane parameters take the values determined 
by naive dimensional analysis.  This implies that our theory does not 
have any hidden fine-tuning and thus is technically natural.}

Finally, $M_S$ in Eqs.~(\ref{eq:Higgs_eff-W-add}) can take essentially 
any value of order the weak scale.  We thus treat it as a free parameter 
in our analysis below.

\subsection{Minimization of the Higgs potential}
\label{subsec:Higgs-min}

We now present examples of parameters for the model of 
section~\ref{sec:theory} that lead to realistic electroweak symmetry 
breaking and acceptable phenomenology.  In Table~\ref{table:points}, 
we list three points (A, B and C) in the parameter space and the 
corresponding values of the fine-tuning parameters defined in 
Eqs.~(\ref{eq:ft-parameter},~\ref{eq:ft-parameter-mod}).  The square 
bracket in the table is defined as $[X]^{n} \equiv {\rm sgn}(X) \cdot 
|X|^{n}$, and the sign convention is such that $\tan\beta \equiv 
\langle H_u \rangle/\langle H_d \rangle > 0$.  We also list the physical 
Higgs-boson mass, some parameters in the Higgs sector, and the soft 
supersymmetry breaking masses for each point.
\begin{table}
\begin{center}
\begin{tabular}{|c|c|c|c|}  \hline 
                             &    A    &    B    &    C   
\\ \hline
 $\lambda$                   &   $0.8$ &   $0.8$ &   $0.8$ \\
 $M_S$                       &   $317$ &     $0$ &     $0$ \\
 $\kappa$                    &     $0$ &   $0.2$ &     $0$ 
\\ \hline
 $[L_S^2]^{1/2}$             &   $-85$ &   $203$ &   $243$ \\
 $[L_S^2 C_S]^{1/3}$         &  $-523$ &  $-464$ &  $-535$ \\
 $[A_\lambda]$               &   $-21$ &    $21$ &    $25$ \\
 $[m_S^2]^{1/2}$             &   $808$ &   $683$ &   $787$ \\
 $[m_{H_u}^2]^{1/2}$         &  $-106$ &  $-102$ &  $-124$ \\
 $[m_{H_d}^2]^{1/2}$         &   $192$ &   $192$ &   $193$ \\
 $k'$   & $8 \times 10^4$ & $7 \times 10^4$ & $7 \times 10^4$ 
\\ \hline
 $\tan\beta$                 &   $1.8$ &   $1.8$ &   $1.7$ \\
 $\mu_{\rm eff}$             &   $156$ &   $159$ &   $192$ \\
 $[(\mu B)_{\rm eff}]^{1/2}$ &  $-132$ &   $167$ &   $208$ \\
 $M_{H, {\rm tree}}$         &   $123$ &   $116$ &   $111$ \\
 $\theta_H$                  & $0.013$ & $0.066$ & $0.081$ \\
 $M_{\rm Higgs}$             &   $140$ &   $134$ &   $130$ 
\\ \hline
 $M_1$                       &   $346$ &   $259$ &   $173$ \\
 $M_2$                       &   $253$ &   $274$ &   $289$ \\
 $M_3$                       &   $305$ &   $307$ &   $328$ \\
 $(m_{\tilde{q}}^2)^{1/2}$   &   $312$ &   $319$ &   $339$ \\
 $(m_{\tilde{u}}^2)^{1/2}$   &   $295$ &   $283$ &   $290$ \\
 $(m_{\tilde{d}}^2)^{1/2}$   &   $271$ &   $269$ &   $285$ \\
 $(m_{\tilde{l}}^2)^{1/2}$   &   $192$ &   $192$ &   $193$ \\
 $(m_{\tilde{e}}^2)^{1/2}$   &   $200$ &   $150$ &   $100$ 
\\ \hline
 $A_t$                       &  $-198$ &  $-195$ &  $-204$ \\
 $M_{\tilde{t}_1}$           &   $221$ &   $214$ &   $212$ \\
 $M_{\tilde{t}_2}$           &   $385$ &   $387$ &   $404$ 
\\ \hline
 $\Delta^{-1}$               &  $19\%$ &  $18\%$ &  $12\%$ \\
 $\tilde{\Delta}^{-1}$       &  $26\%$ &  $24\%$ &  $22\%$ 
\\ \hline
\end{tabular}
\end{center}
\caption{Values for the parameters of the model for three sample points, 
 A, B and C.  The resulting soft supersymmetry breaking parameters as 
 well as the quantities in the Higgs sector are also listed.  Here, 
 $[X]^{n} \equiv {\rm sgn}(X) \cdot |X|^{n}$, and all masses are 
 given in units of GeV.}
\label{table:points}
\end{table}

The procedure to obtain the numbers in the table is as follows.  (i) We 
first choose the rescaled AdS curvature scale, $k'$, and choose $\lambda$, 
$M_S$, $\kappa$, $L_S^2$, $L_S^2 C_S$, $A_\lambda$ and $m_S^2$ as free 
parameters, which are roughly within the bounds discussed above.  (ii) 
The values of $m_{H_u}^2$ and $m_{H_d}^2$ are also chosen arbitrarily. 
This gives $\langle S \rangle$, $\langle H_u \rangle$ and $\langle 
H_d \rangle$ through the minimization of the Higgs potential 
of Eq.~(\ref{eq:Higgs-Pot}) supplemented by the terms in 
Eqs.~(\ref{eq:Higgs_eff-W-add},~\ref{eq:Higgs_eff-L-add}).  The values 
of $\lambda$, $M_S$, $\kappa$, $L_S^2$, $L_S^2 C_S$, $A_\lambda$, $m_S^2$, 
$m_{H_u}^2$ and $m_{H_d}^2$ should satisfy one constraint $v \equiv 
(\langle H_u \rangle^2 + \langle H_d \rangle^2)^{1/2} \simeq 174~{\rm GeV}$. 
However, the correct values for these parameters are easily obtained by 
starting from arbitrary values, and then rescaling all the parameters 
according to their dimensions such that they give $v \simeq 174~{\rm GeV}$. 
(iii) At this point, we obtain three numbers in the table: $\tan\beta 
= \langle H_u \rangle/\langle H_d \rangle$, $\mu_{\rm eff} \equiv 
\lambda \langle S \rangle$ and $(\mu B)_{\rm eff} \equiv \lambda(L_S^2 
+ A_\lambda \langle S \rangle - \lambda \sin\beta \cos\beta\, v^2)$. 
We also obtain the tree-level Higgs-boson mass $M_{H, {\rm tree}}$ by 
diagonalizing the $3 \times 3$ scalar mass-squared matrix in the space 
of $\{ {\rm Re}S,\, {\rm Re}H_u^0,\, {\rm Re}H_d^0 \}$ and finding the 
lightest eigenvalue.  The corresponding eigenvector in this space is 
parameterized as $\{ \sin\theta_H,\, \cos\theta_H \cos\varphi_H,\, 
\cos\theta_H \sin\varphi_H \}$, and we also list the value of $\theta_H$, 
i.e. the amount of a singlet component in the lightest Higgs boson mass.
(iv) We then choose the right-handed selectron mass $m_{\tilde{e}}^2$ 
as an input parameter, which satisfies the experimental bound of 
$m_{\tilde{e}}^2 \simgt (100~{\rm GeV})^2$.  This determines the $U(1)_Y$ 
component of the supersymmetry breaking parameters $M_{{\rm SUSY},1}$, 
defined by $M_{{\rm SUSY},a} \equiv (\zeta_a F_Z/M_*)(k'/k)$ ($a=1,2,3$), 
through Eq.~(\ref{eq:scalar-masses-2}).  Similarly, $m_{H_d}^2$ determines 
the $SU(2)_L$ component, $M_{{\rm SUSY},2}$.  Namely,
\begin{equation}
  \left\{ \begin{array}{lll} 
    m_{\tilde{e}}^2\!\!\! &=& 
      \!\! m_{\tilde{f}}^2 \Bigl( C_1^{\tilde{f}}=\frac{3}{5}, 
        C_2^{\tilde{f}}=0, C_3^{\tilde{f}}=0 \Bigr), \\
    m_{H_d}^2\!\!\! &=& 
      \!\! m_{\tilde{f}}^2 \Bigl( C_1^{\tilde{f}}=\frac{3}{20}, 
        C_2^{\tilde{f}}=\frac{3}{4}, C_3^{\tilde{f}}=0 \Bigr), 
  \end{array} \right.
  \longrightarrow \:\: M_{{\rm SUSY},1},\,\, M_{{\rm SUSY},2}.
\end{equation}
Here, $m_{\tilde{f}}^2$ is given in Eq.~(\ref{eq:scalar-masses-2}), and for 
the theory with boundary condition $SU(5)$ breaking the value of $(g_B^2 k)$ 
is determined by the condition that the 321 gauge couplings become strong at 
the scale $k$ in the 4D picture: $(g_B^2 k) = 8\pi^2/b^{\rm DSB} \simeq 16$. 
(v) In our theory the difference between $m_{H_u}^2$ and $m_{H_d}^2$ must 
come essentially from the top Yukawa contribution, so approximately 
\begin{equation}
  m_{H_u}^2 = m_{H_d}^2 -\frac{3y_t^2}{8\pi^2} 
    \Biggl\{ m_{\tilde{q}}^2 \ln\Biggl( \frac{k'}{m_{\tilde{q}}} \Biggr) 
    + m_{\tilde{u}}^2 \ln\Biggl( \frac{k'}{m_{\tilde{u}}} \Biggr) \Biggr\}.
\label{eq:mHu-mHd-rel}
\end{equation}
Here, we have simply cut off the UV-divergent logarithm arising in the 
4D one-loop calculation by the rescaled AdS scale $k'$, which approximates 
the full 5D finite computation reasonably well.%
\footnote{In many 5D calculations, the effective cutoff scales of 
4D divergent integrals are smaller than the KK gauge-boson mass of 
$m_{\rm KK} \simeq (3\pi/4)k' \simeq 2.4 k'$.}
The top Yukawa coupling is given by $m_t/v\sin\beta$.  This equation 
fixes $M_{{\rm SUSY},3}$, the only parameter in Eq.~(\ref{eq:mHu-mHd-rel})
still undetermined:
\begin{equation}
  \left\{ \begin{array}{lll} 
    m_{\tilde{q}}^2\!\!\! &=& 
      \!\! m_{\tilde{f}}^2 \Bigl( C_1^{\tilde{f}}=\frac{1}{60}, 
        C_2^{\tilde{f}}=\frac{3}{4}, C_3^{\tilde{f}}=\frac{4}{3} \Bigr), \\
    m_{\tilde{u}}^2\!\!\! &=& 
      \!\! m_{\tilde{f}}^2 \Bigl( C_1^{\tilde{f}}=\frac{4}{15}, 
        C_2^{\tilde{f}}=0, C_3^{\tilde{f}}=\frac{4}{3} \Bigr), 
  \end{array} \right.
  \longrightarrow \:\: M_{{\rm SUSY},3}.
\end{equation}
Therefore, from Eqs.~(\ref{eq:gaugino-masses-2},~\ref{eq:scalar-masses-2}), 
We obtain the soft supersymmetry breaking masses for the gauginos $M_1$, 
$M_2$ and $M_3$, and for the scalars $m_{\tilde{q}}^2$, $m_{\tilde{u}}^2$, 
$m_{\tilde{d}}^2$, $m_{\tilde{l}}^2$ and $m_{\tilde{e}}^2$.  (vi) The 
physical top-squark masses, $M_{\tilde{t}_1}$ and $M_{\tilde{t}_2}$ 
($M_{\tilde{t}_2} > M_{\tilde{t}_1}$), are given by diagonalizing the 
mass-squared matrix
\begin{equation}
  M_{\tilde{t}}^2 = 
  \pmatrix{
     m_{\tilde{q}_3}^2 + m_t^2 
       + \Bigl( \frac{1}{2} - \frac{2}{3}\sin^2\!\theta_w \Bigr) 
         \cos 2\beta\, M_Z^2 & 
     y_t v (A_t \sin\beta - \mu_{\rm eff} \cos\beta) \cr 
     y_t v (A_t \sin\beta - \mu_{\rm eff} \cos\beta) & 
     m_{\tilde{u}_3}^2 + m_t^2 
       + \frac{2}{3}\sin^2\!\theta_w\, \cos 2\beta\, M_Z^2 \cr 
  },
\label{eq:stop-massmat}
\end{equation}
where $\theta_w$ is the Weinberg angle.  Here, the soft supersymmetry 
breaking masses for the third-generation squarks, $m_{\tilde{q}_3}^2$ 
and $m_{\tilde{u}_3}^2$, are given by
\begin{eqnarray}
  && m_{\tilde{q}_3}^2 \:\simeq\: m_{\tilde{q}}^2 
    -\frac{1}{3} (m_{H_d}^2 - m_{H_u}^2),
\\
  && m_{\tilde{u}_3}^2 \:\simeq\: m_{\tilde{u}}^2 
    -\frac{2}{3} (m_{H_d}^2 - m_{H_u}^2),
\end{eqnarray}
while the scalar trilinear coupling, $A_t$, is given by
\begin{equation}
  A_t \:\simeq\: -\frac{1}{16\pi^2} 
    \Biggl\{ \frac{32}{3}g_3^2 M_3 \ln\Biggl( \frac{k'}{M_3} \Biggr)
    + 6 g_2^2 M_2 \ln\Biggl( \frac{k'}{M_2} \Biggr)
    + \frac{26}{15}g_1^2 M_1 \ln\Biggl( \frac{k'}{M_1} \Biggr) \Biggr\}.
\end{equation}
(vii) With the numbers obtained so far, we can calculate the radiative 
correction to the Higgs potential arising from the top Yukawa coupling, 
which is the dominant source of radiative corrections.  It gives a
correction to the Higgs potential 
\begin{equation}
  \delta V = \frac{\lambda_{H, {\rm top}}}{2} |H_u|^2,
\end{equation}
where $\lambda_{H, {\rm top}}$ is given, at one loop, by~\cite{Lopez:1991aw}
\begin{eqnarray}
  \lambda_{H, {\rm top}} &\simeq& \frac{3 y_t^4}{16\pi^2} 
    \Biggl[ \ln\left(\frac{M_{\tilde{t}_2}^2 M_{\tilde{t}_1}^2}{m_t^4}\right)
\nonumber\\
    && + \left(\frac{M_{\tilde{t}_2}^2-M_{\tilde{t}_1}^2}{4m_t^2} 
      \sin^2\! 2\theta_{\tilde{t}} \right)^2 
      \!\! f(M_{\tilde{t}_2}^2,M_{\tilde{t}_1}^2)
    + \frac{M_{\tilde{t}_2}^2-M_{\tilde{t}_1}^2}{2m_t^2} 
      \sin^2\! 2\theta_{\tilde{t}} 
      \ln\left(\frac{M_{\tilde{t}_2}^2}{M_{\tilde{t}_1}^2}\right) \Biggr],
\end{eqnarray}
where $f(x,y) \equiv 2-((x+y)/(x-y))\ln(x/y)$, and $\theta_{\tilde{t}}$ 
is the mixing angle for the top squarks needed to go from the basis of 
Eq.~(\ref{eq:stop-massmat}) to the mass eigenbasis.  With this correction 
added to the Higgs potential, we can now iterate the procedure from (i) 
to (vii) until it converges.  All the values listed in the table are then 
corrected by this iteration procedure, except $M_{H, {\rm tree}}$ which is 
by definition a tree-level quantity.  We find that the convergence is rather 
quick, and the corrections are not so large.  (viii) Finally, the fine-tuning 
parameter $\Delta^{-1}$ is obtained by slightly varying the fundamental 
parameters of the theory and measuring the response of $M_Z$ under that 
variation.  This parameter is defined in Eq.~(\ref{eq:ft-parameter}), 
and the fundamental parameters $a_i$ are taken as $k'$, $\lambda$, $M_S$, 
$\kappa$, $L_S^2$, $L_S^2 C_S$, $A_\lambda$, $m_S^2$, $M_{{\rm SUSY},1}$, 
$M_{{\rm SUSY},2}$ and $M_{{\rm SUSY},3}$.  In most of the parameter space, 
the dominant contribution comes from $a_i = M_{{\rm SUSY},3}$, $L_S^2 C_S$ 
or $m_S^2$.  In calculating $\Delta^{-1}$ we do not include the gauge and 
Yukawa couplings in $a_i$, because $M_Z^2$ has large generic sensitivities 
to these parameters.  These parameters, however, are included in $a_i$ when 
we calculate $\tilde{\Delta}^{-1}$ defined in Eq.~(\ref{eq:ft-parameter-mod}). 
The parameters $\eta_i$ are estimated naively from the dependence of 
$M_Z^2$ on each parameter in ``generic'' parameter regions.  For the 
relevant parameters this gives: $\eta_i = 1/2$ for $\{ \lambda, L_S^2 C_S, 
m_S^2, M_{{\rm SUSY},i}, y_t \}$, $\eta_i \simlt 1/3$ for $\{M_S, L_S^2 \}$ 
and $\eta_i = 1/4$ for $\{ g_1, g_2, g_3 \}$.  The dominant contribution 
to $\tilde{\Delta}^{-1}$ comes from $a_i = y_t$.  The parameters 
$\Delta^{-1}$ and $\tilde{\Delta}^{-1}$ provide rough measures for 
fine-tuning required in our theory.

The numbers in Table~\ref{table:points} are subject to errors at the
$10\%$ level for the soft superparticle masses.  The error could be somewhat 
larger, at the $20\%$ level for $M_3$, due to the strong sensitivity 
of $g_3$ to the renormalization scale.  The error for the Higgs-boson 
mass is expected to be at the level of a few GeV.  Note that, in contrast 
to the MSSM case, the two-loop radiative correction to the Higgs potential 
is not very large in our theory.  This is because the top squarks are 
rather light, $\approx 300~{\rm GeV}$, so that the logarithm appearing in 
the radiative correction, $\ln(M_{\tilde{t}_2}^2 M_{\tilde{t}_1}^2/m_t^4)$, 
is not so large.  Comparing with the full two-loop calculation of the 
radiative correction~\cite{Heinemeyer:1998yj}, we estimate that the 
two-loop contribution to the physical Higgs boson mass (overestimate 
of $M_{\rm Higgs}$ in Table~\ref{table:points}) is about $4~{\rm GeV}$. 

\begin{table}
\begin{center}
\begin{tabular}{|c|c|c|c|}  \hline 
                  &    A    &    B    &    C    
\\ \hline
 $\tilde{g}$      &  $305$  &  $307$  &  $328$  
\\ \hline
 $\chi^{\pm}_1$   &  $115$  &  $121$  &  $150$  \\
 $\chi^{\pm}_2$   &  $297$  &  $314$  &  $332$  
\\ \hline
 $\chi^0_1$       &  $103$  &   $88$  &   $56$  \\
 $\chi^0_2$       &  $193$  &  $162$  &  $132$  \\
 $\chi^0_3$       &  $288$  &  $221$  &  $200$  \\
 $\chi^0_4$       &  $353$  &  $262$  &  $263$  \\
 $\chi^0_5$       &  $365$  &  $321$  &  $336$  
\\ \hline
 $H^0_1$          &  $140$  &  $134$  &  $130$  \\
 $H^0_2$          &  $298$  &  $304$  &  $332$  \\
 $H^0_3$          &  $872$  &  $718$  &  $802$  
\\ \hline
 $P^0_1$          &  $305$  &  $315$  &  $343$  \\
 $P^0_2$          &  $888$  &  $687$  &  $799$  
\\ \hline
 $H^{\pm}$        &  $288$  &  $293$  &  $323$  
\\ \hline
 $\tilde{u}_L$    &  $309$  &  $317$  &  $337$  \\
 $\tilde{u}_R$    &  $294$  &  $281$  &  $289$  \\
 $\tilde{d}_L$    &  $315$  &  $322$  &  $341$  \\
 $\tilde{d}_R$    &  $272$  &  $270$  &  $285$  \\
 $\tilde{e}_L$    &  $195$  &  $195$  &  $196$  \\
 $\tilde{e}_R$    &  $203$  &  $153$  &  $105$  \\
 $\tilde{\nu}_L$  &  $186$  &  $187$  &  $188$  
\\ \hline
 $\tilde{t}_1$    &  $221$  &  $214$  &  $212$  \\
 $\tilde{t}_2$    &  $385$  &  $387$  &  $404$  
\\ \hline
\end{tabular}
\end{center}
\caption{The masses for the superparticles and the Higgs bosons for 
 three sample points A, B and C given in Table~\ref{table:points}. 
 All masses are given in units of GeV.}
\label{table:spectra}
\end{table}
The superparticle masses in the parameter points A, B and C, are listed 
in Table~\ref{table:spectra}, in which we present the mass eigenvalues for 
the 2 charginos, $\chi^{\pm}_{1,2}$, 5 neutralinos, $\chi^0_{1,2,3,4,5}$, 
3 neutral scalar Higgs bosons, $H^0_{1,2,3}$, 2 neutral pseudo-scalar Higgs 
bosons, $P^0_{1,2}$, and charged Higgs bosons, $H^{\pm}$.  We also list 
the masses for the scalars, $\tilde{u}_L$, $\tilde{u}_R$, $\tilde{d}_L$, 
$\tilde{d}_R$, $\tilde{e}_L$, $\tilde{e}_R$ and $\tilde{\nu}_L$, which 
include the $D$-term contributions.  The masses for the 2 top squarks 
are listed separately, as they split from the other superparticles 
by non-negligible amounts.  All three points evade phenomenological 
constraints such as direct collider searches for the superparticles 
(the issue of evading the constraints on neutralinos for points B and C 
will be discussed in section~\ref{subsec:neutralino}).  As expected, 
we find that the superparticle masses are rather light and close to 
experimental bounds.  We also see from the table that the superparticle 
spectrum is quite different from one characteristic of conventional 
unified theories.  In particular, the hierarchy among the three gaugino 
masses is typically much smaller than the one arising from the unified 
gaugino mass relation $M_1 : M_2 : M_3 \simeq g_1^2 : g_2^2 : g_3^2$. 

Because of the rather small masses for the charged Higgs boson, the $b 
\rightarrow s \gamma$ process could potentially give strong constraints 
on our theory.  The branching ratio for this process is measured fairly 
accurately: ${\rm Br}(b \rightarrow s \gamma) = (3.3 \pm 0.4) \times 
10^{-4}$ at the $1\sigma$ level~\cite{Eidelman:2004wy}, which agrees well 
with the standard model prediction.  The contribution from the charged 
Higgs boson always interferes constructively with the standard-model 
contribution.  For a charged Higgs boson mass in the range $\approx 
(250\!\sim\!350)~{\rm GeV}$, which covers the values obtained in the 
points presented in Table~\ref{table:spectra}, the next-to-leading order 
QCD calculation gives the sum of the contributions from the standard model 
and the charged Higgs boson at the level ${\rm Br}(b \rightarrow s \gamma) 
\simeq (4\!\sim\!5) \times 10^{-4}$~\cite{Borzumati:1998tg}, which is 
somewhat larger than the observed value.  There is, however, also a 
contribution from chargino loops.  In our theory, this contribution 
interferes destructively (constructively) with the standard-model one 
if the sign of $\mu_{\rm eff}$ is positive (negative).  This, therefore, 
prefers the positive sign for $\mu_{\rm eff}$ (and thus certain signs 
for the fundamental parameters).  With $\mu_{\rm eff} > 0$, we find that 
the prediction for ${\rm Br}(b \rightarrow s \gamma)$ in our theory can 
naturally be consistent with experimental data.  The contributions from 
neutralino and gluino loops are negligible.

Finally, we emphasize that it is significant that our theory reduces 
the fine-tuning down to the level $\Delta^{-1} = O(10\!\sim\!20\%)$ 
($\tilde{\Delta}^{-1} = O(20\!\sim\!30\%)$) as presented in the table. 
As we have seen in section~\ref{sec:sources}, most existing supersymmetry 
breaking scenarios leads to a fine-tuning at the $3\%$ level or even worse. 
We have also seen that even with rather general superparticle masses, 
the fine-tuning is still worse than about $5\%$ in the MSSM.  Our theory 
does not need such an accurate cancellation among different parameters. 
In fact, we expect that the level of tuning (given by the sizes of 
$\Delta^{-1}$ and $\tilde{\Delta}^{-1}$) obtained in Table~\ref{table:points} 
is close to the best we can attain in theories that accommodate the MSSM 
sector in a perturbative way.

\section{Phenomenological Issues}
\label{sec:pheno}

In this section we discuss some of the phenomenological issues in our 
theory, focusing on neutralino phenomenology and pedestrian 
dark matter in particular. 

\subsection{Neutralino phenomenology}
\label{subsec:neutralino}

As we saw in the previous subsection, our theory generically predicts 
light superparticles, and the neutralinos can be particularly light. 
We here consider the phenomenology of the neutralino sector.

Let us start with the point~A in Tables~\ref{table:points} and 
\ref{table:spectra}.  For this point, the mass of the lightest neutralino 
is larger than half of the LEP~II center-of-mass energy $\sqrt{s} \simeq 
200~{\rm GeV}$, so there is no constraint from direct searches.  On 
the other hand, for the sample points B and C, the mass of the lightest 
neutralino is around $90~{\rm GeV}$ and $60~{\rm GeV}$, respectively 
(see Table~\ref{table:spectra}).  Such light neutralinos could be 
dangerous, as they contain non-negligible Higgsino components and are 
easily produced at $e^+e^-$ colliders through $s$-channel $Z$ exchanges. 
Whether these light neutralinos evade experimental constraints or not, 
then, depends on their compositions and decay channels. 

The masses and compositions of the lightest neutralino, $\chi^0_1$, for 
three sample points A, B and C are given in Table~\ref{table:chi1}. 
\begin{table}
\begin{center}
\begin{tabular}{|c|c|c|}  \hline 
    &  $m_{\chi^0_1}$  &  $\{ c_{1,\tilde{B}},\, c_{1,\tilde{W}_3},\, 
       c_{1,\tilde{H}_d},\, c_{1,\tilde{H}_u},\, c_{1,\tilde{S}} \}$  
\\ \hline
 A & $103~{\rm GeV}$ & $\{ -0.15,\,  0.45,\, -0.60,\,  0.63,\, -0.14 \}$ \\
 B & $88~{\rm GeV}$  & $\{  0.20,\, -0.33,\,  0.36,\, -0.69,\,  0.50 \}$ \\
 C & $56~{\rm GeV}$  & $\{  0.16,\, -0.15,\, -0.060,\, -0.54,\, 0.81 \}$ 
\\ \hline
\end{tabular}
\end{center}
\caption{The masses and compositions of the lightest neutralino $\chi^0_1$ 
 for the three sample points A, B and C given in Table~\ref{table:points}.}
\label{table:chi1}
\end{table}
Here, the coefficients $c_{1,\tilde{B}}, c_{1,\tilde{W}_3}, c_{1,\tilde{H}_d}, 
c_{1,\tilde{H}_u}$ and $c_{1,\tilde{S}}$ are defined through the relations 
between the mass and gauge eigenstates:
\begin{equation}
  \chi_i = c_{i,\tilde{B}} \tilde{B} + c_{i,\tilde{W}_3} \tilde{W}_3 
    + c_{i,\tilde{H}_d} \tilde{H}_d + c_{i,\tilde{H}_u} \tilde{H}_u 
    + c_{i,\tilde{S}} \tilde{S},
\end{equation}
where $i=1,\cdots,5$, and $c$'s are normalized as $c_{i,\tilde{B}}^2 
+ c_{i,\tilde{W}_3}^2 + c_{i,\tilde{H}_d}^2 + c_{i,\tilde{H}_u}^2 
+ c_{i,\tilde{S}}^2 = 1$.  From the table, one sees that the Higgsino 
components in $\chi^0_1$ are in fact non-negligible.  This is because 
the sizes of $\mu_{\rm eff}$ and $M_1$ are comparable for these parameter 
points (see Table~\ref{table:points}).  One also finds that $\chi^0_1$ 
generically contains non-negligible amounts of the fermionic component 
of $S$. 

As we already discussed, the point~A evades direct search constraints 
regardless of the decay of $\chi^0_1$, but this is not true for the 
points B and C.  What are the decay modes of $\chi^0_1$?  Since $\chi^0_1$ 
is $R$-parity odd, its decay products must include an $R$-parity odd 
particle with mass smaller than $m_{\chi^0_1}$.  An obvious candidate 
for such particle is the gravitino, which is generically very light, 
with mass $m_{3/2}$ given by 
\begin{equation}
  m_{3/2} \simeq \frac{F_Z^{\prime\, 2}}{M_{\rm Pl}} 
    \simeq (0.1\!\sim\!10)~{\rm eV},
\end{equation}
where $F'_Z \equiv F_Z e^{-2\pi kR} \approx \{(10\!\sim\!100)~{\rm 
TeV}\}^2$ is the rescaled supersymmetry breaking scale (see e.g. 
Eq.~(\ref{eq:gaugino-masses-2})).  If this were the dominant decay 
channel, these points would be excluded, because then a significant 
fraction of the decay of $\chi^0_1$ would go to the gravitino and some 
visible particles, such as the photon, and such a signal would have 
already been observed at LEP~II.  In our theory, however, $\chi^0_1$ can 
also decay into a pair of the pedestrian fields, depending on the masses 
of these fields.  Therefore, it is not obvious that these points are 
excluded by the present experimental data. 

To illustrate this point, we introduce a Planck-brane pedestrian field 
$P''$ which has sufficiently small mixing with $P$, i.e. a sufficiently 
small coefficient for the superpotential term of the form $W = S P P''$. 
The superpotential for this field is then given by 
\begin{equation}
  W \simeq \frac{h''}{2} S P''^2 + \frac{M_{P''}}{2} P''^2.
\label{eq:S-P''}
\end{equation}
In order for $P''$ not to have a VEV, we need $h'' \langle F_S \rangle \simlt 
|M_{P''}|^2$.  The lightest neutralino can then decay into fermionic and 
scalar components of $P''$, if $|M_{P''}| \simlt m_{\chi^0_1}/2$.  Since 
$\langle F_S \rangle \simeq (200~{\rm GeV})^2$ in generic parameter regions, 
this implies that $h'' \simlt 0.05$ and $0.02$ for the points B and C, 
respectively.  Such small couplings do not affect the RG analysis of 
section~\ref{subsec:para-Higgs}. 

With the coupling of $S$ to $P''$ in Eq.~(\ref{eq:S-P''}), the constraints 
from direct searches are evaded.  The constraints from the $Z$-pole data 
at LEP~I are also easily evaded due to phase-space suppression and the 
fact that the dominant decay of $\chi^0_1$ is invisible.  At LEP~II with 
$\sqrt{s} \simeq 200~{\rm GeV}$, two on-shell $\chi^0_1$'s can be produced. 
However, since the branching ratio of $\chi^0_1$ decay into visible 
particles is smaller than about $10^{-4}$, this process evades detection 
and thus the corresponding parameter region is not excluded. 

\begin{table}
\begin{center}
\begin{tabular}{|c|c|c|}  \hline 
    &  $m_{\chi^0_2}$  &  $\{ c_{2,\tilde{B}},\, c_{2,\tilde{W}_3},\, 
       c_{2,\tilde{H}_d},\, c_{2,\tilde{H}_u},\, c_{2,\tilde{S}} \}$  
\\ \hline
 A & $193~{\rm GeV}$ & $\{ 0.021,\, -0.046,\, -0.69,\, -0.67,\, -0.25 \}$ \\
 B & $162~{\rm GeV}$ & $\{ -0.16,\,  0.26,\,  -0.57,\, 0.086,\,  0.76 \}$ \\
 C & $132~{\rm GeV}$ & $\{ -0.61,\,  0.29,\,  -0.57,\,  0.31,\,  0.34 \}$ 
\\ \hline
\end{tabular}
\end{center}
\caption{The masses and compositions of the next-to-lightest 
 neutralino $\chi^0_2$ for three sample points A, B and C given 
 in Table~\ref{table:points}.}
\label{table:chi2}
\end{table}
For parameter point~C, there is a potential danger coming from 
$\chi^0_1$-$\chi^0_2$ associated production.  While the production rate 
receives a suppression of about $0.03$ (see the compositions of $\chi^0_2$ 
in Table~\ref{table:chi2}), this still requires the branching ratio of 
visible $\chi^0_2$ decays to be smaller than about $3\%$, if the production 
occurs with full strength.  The precision of our calculation, however, 
allows errors of order $O(10\%)$ for the masses, so we cannot conclude 
that the production actually occurs for this parameter point.  In any event, 
since $\chi^0_2$ dominantly decays into $\chi^0_1$ and visible particles, 
this process constrains the sum of the masses of $\chi^0_1$ and $\chi^0_2$ 
to be larger than about $200~{\rm GeV}$ so that their production at 
LEP~II is suppressed.

\subsection{Pedestrian dark matter}
\label{subsec:pedestrian-DM}

Because of the unbroken $P$ parity, the lightest pedestrian field is 
absolutely stable.  It may therefore constitute the dark matter of the 
universe, depending on the parameters of the model.  Here we discuss this 
issue for the simplest case of a single Planck-brane pedestrian field, 
discussed in section~\ref{subsec:warped} (without additional fields $P''$).

Since the bulk pedestrian fields receive a large supersymmetric mass of 
order $10~{\rm TeV}$ on the TeV brane, these fields are much heavier than 
the Planck-brane pedestrian multiplet $P'$.  The lightest pedestrian field 
is thus a component of $P'$.  In the model of section~\ref{subsec:warped}, 
the scalar component of $P'$ obtains a soft supersymmetry-breaking mass 
of about $(600\!\sim\!800)~{\rm GeV}$ (see Eq.~(\ref{eq:mS2-mPp2-5D})) 
while the mass of the fermionic component is given by $h \langle S \rangle 
+ M_{P'}$.  Since $\langle S \rangle \simeq 200~{\rm GeV}$ and $h \simlt 
0.4$ (from the RG analysis; see section~\ref{subsec:para-Higgs}), we 
expect that the lightest pedestrian field is the fermionic component 
of $P'$, $\psi_{P'}$, for a wide range of $M_{P'}$, which is essentially 
a free parameter of the theory. 

In the minimal case considered here, the annihilation of $\psi_{P'}$ 
occurs through $s$-channel exchange of the $S$ scalar, with mass 
about $(700\!\sim\!800)~{\rm GeV}$, or through $t$-channel exchange 
of $\psi_{P'}$ using the mixing between the $S$ and Higgs scalars. 
Let us assume, for simplicity, that the mass of $\psi_{P'}$ is larger 
than the Higgsino mass so that the annihilation into two Higgsinos are 
kinematically allowed.  In this case, the dominant contribution to the 
annihilation comes from the diagram with the $s$-channel $S$-scalar 
exchange, giving the thermally averaged cross section of order $\langle 
\sigma v \rangle \simeq ((\lambda h)^2/8\pi)(m_{\psi_{P'}}^2/m_S^4)$. 
This can easily give the correct abundance for dark matter in our 
generic parameter region, $\lambda \simeq 0.8$, $h \simlt 0.4$ and 
$m_S \simeq (700\!\sim\!800)~{\rm GeV}$ with $m_{\psi_{P'}}$ essentially 
a free parameter in a range $m_{\psi_{P'}} < (600\!\sim\!800)~{\rm GeV}$. 
Note that $\psi_{P'}$ annihilation into the two Higgsinos is not 
subject to the $p$-wave suppression, so that we can naturally obtain 
the correct dark matter abundance with the masses of $O(100~{\rm 
GeV}\!\sim\!1~{\rm TeV})$.  This implies that the relic abundance does 
not change much even in the case that the pedestrian dark matter is 
a Dirac fermion.  The annihilation rate, however, can be enhanced for 
$M_{\psi_{P'}} \simeq m_S/2 \simeq 400~{\rm GeV}$ due to the $s$-channel 
$S$-scalar pole, resulting in a significant reduction of the relic 
abundance.

\section{Purely 4D Realizations}
\label{sec:alternative}

In this section we present an outline for constructing purely 4D 
theories with reduced fine-tuning.  First, we make the logarithm in 
Eq.~(\ref{eq:corr-Higgs}) smaller by requiring a low mediation scale of 
supersymmetry breaking.  In particular, we consider theories in which the 
fundamental scale of supersymmetry breaking is of order $100~{\rm TeV}$. 
Such theories, with supersymmetry breaking mediated by standard-model 
gauge interactions, were constructed, for example, in 
Refs.~\cite{Izawa:1997gs,Izawa:2005yf}.  Here, we adopt the basic 
construction of~\cite{Izawa:2005yf} to illustrate our point. 

The DSB sector consists of an $S\!P(2)$ gauge theory with 6 chiral 
superfields $\tilde{Q}_i$ ($i=1,\cdots,6$) in the fundamental 4-dimensional 
representation, together with 15 singlets $Z^a$ ($a=1,\cdots,14$) and $Z$. 
With the tree-level superpotential $W = \lambda' Z^a (\tilde{Q}\tilde{Q})_a 
+ \lambda Z (\tilde{Q}\tilde{Q})$, where $(\tilde{Q}\tilde{Q})_a$ denotes 
a flavor 14-plet of $S\!P(3)_{\rm flavor}$ unbroken after the inclusion 
of the superpotential, supersymmetry is broken.  For a certain parameter 
region, the supersymmetry breaking VEV is given by $F_Z \simeq \lambda 
\Lambda^2$, where $\Lambda$ is the dynamical scale of $S\!P(2)$ gauge 
interactions.  We assume throughout that the $Z$ field does not have 
a VEV, $\langle Z \rangle = 0$.  In fact, this point is (at least) a 
local minimum of the potential~\cite{Chacko:1998si}.

Supersymmetry breaking is mediated to the SSM sector both by vector-like 
mediator fields, which are charged under both 321 and $S\!P(2)$ gauge 
interactions, and by vector-like messenger fields, which are charged only 
under 321.  In particular, we introduce mediator fields ${\cal D}({\bf 3}^*, 
{\bf 1})_{1/3}$, $\bar{\cal D}({\bf 3}, {\bf 1})_{-1/3}$, ${\cal L}({\bf 1}, 
{\bf 2})_{-1/2}$ and $\bar{\cal L}({\bf 1}, {\bf 2})_{1/2}$ that are in the 
fundamental representation of the supersymmetry-breaking $S\!P(2)$ gauge 
group.  Here, the numbers in parentheses denote quantum numbers under 321. 
We also introduce messenger fields ${\cal D}'({\bf 3}^*, {\bf 1})_{1/3}$, 
$\bar{\cal D}'({\bf 3}, {\bf 1})_{-1/3}$, ${\cal L}'({\bf 1}, {\bf 2})_{-1/2}$ 
and $\bar{\cal L}'({\bf 1}, {\bf 2})_{1/2}$ that are singlet under the 
$S\!P(2)$ gauge group.  The superpotential for these fields is given by
\begin{equation}
  W = m_{\cal D} {\cal D} \bar{\cal D} + m_{\cal L} {\cal L} \bar{\cal L}
    + (m'_{\cal D} + k_{\cal D} Z) {\cal D}' \bar{\cal D}' 
    + (m'_{\cal L} + k_{\cal L} Z) {\cal L}' \bar{\cal L}',
\label{eq:mediators}
\end{equation}
where $m_{\cal D}$, $m_{\cal L}$, $m'_{\cal D}$, $m'_{\cal L}$ are mass 
parameters that are assumed to be of order $4\pi \Lambda$.%
\footnote{These mass parameters may be generated by the nonperturbative 
dynamics of some new gauge interaction.  The dynamical scale of this 
new gauge theory may be related to that of the supersymmetry-breaking 
$S\!P(2)$ gauge interaction if there are massive fields that are charged 
under both gauge groups and which decouple at a scale somewhat larger 
than $4\pi \Lambda$. A simple example of such behavior arises if the 
couplings of both gauge groups are rather large but their beta functions 
small above the scale where the massive fields decouple.}
An important point is that we take $m_{\cal D} \neq m_{\cal L}$ 
($m_{\cal D} > m_{\cal L}$) and $m'_{\cal D}/k_{\cal D} \neq 
m'_{\cal L}/k_{\cal L}$ ($m'_{\cal D}/k_{\cal D} > m'_{\cal L}/k_{\cal L}$), 
so that the mediator/messenger sector does not respect an approximate 
$SU(5)$ symmetry.  As we will see, this breaks the unwanted mass 
relation of Eq.~(\ref{eq:te-ratio}).  The successful prediction for gauge 
coupling unification, on the other hand, is preserved because the mediator 
and messenger fields fill complete multiplets of $SU(5)$.  Breaking the 
$SU(5)$ structure in the messenger sector of gauge mediation models was 
also considered in~\cite{Agashe:1997kn} to reduce fine-tuning.  Note that 
our choice of mediator fields corresponds to 4 pairs of ${\bf 5} + {\bf 5}^*$ 
under $SU(5)$, so that we have 5 pairs of ${\bf 5} + {\bf 5}^*$ in total. 
This makes 321 strongly coupled, $g_a \sim 4\pi$, at the unification scale, 
as in the models of section~\ref{sec:theory}. 

The masses of the MSSM gauginos are generated when the messenger fields are 
integrated out at the scale $4\pi \Lambda$.  The 321 gaugino masses, $M_a$, 
are given by
\begin{equation}
  M_1 \simeq \frac{g_1^2}{16\pi^2} 
    \left( \frac{2}{5}\frac{k_{\cal D} F_Z}{m'_{\cal D}} 
    + \frac{3}{5}\frac{k_{\cal L} F_Z}{m'_{\cal L}} \right),
\quad
  M_2 \simeq \frac{g_2^2}{16\pi^2} \frac{k_{\cal L} F_Z}{m'_{\cal L}},
\quad
  M_3 \simeq \frac{g_3^2}{16\pi^2} \frac{k_{\cal D} F_Z}{m'_{\cal D}}.
\label{eq:gaugino-4D}
\end{equation}
The sfermion masses, on the other hand, receive contributions both from 
mediator and messenger fields and are given by
\begin{equation}
  m_{\tilde{f}}^2 \simeq 2 \sum_{a=1,2,3} 
    \frac{g_a^4 C_a^{\tilde{f}}}{(16\pi^2)^2} 
    \left( \frac{2 |\lambda|^2 |F_Z|^2}{\tilde{m}_a^2} 
  + \tilde{m}_a'^2 \right),
\label{eq:sfermion-4D}
\end{equation}
where $\tilde{f} = \tilde{q}, \tilde{u}, \tilde{d}, \tilde{l}, \tilde{e}$, 
and $C_a^{\tilde{f}}$ are given by $(C_1^{\tilde{f}}, C_2^{\tilde{f}}, 
C_3^{\tilde{f}}) = (1/60,3/4,4/3)$, $(4/15,0,4/3)$, $(1/15,0,4/3)$, 
$(3/20,3/4,0)$ and $(3/5,0,0)$ for $\tilde{f} = \tilde{q}, \tilde{u}, 
\tilde{d}, \tilde{l}$ and $\tilde{e}$, respectively.  The mass parameters 
$\tilde{m}_a^2$ and $\tilde{m}_a'^2$ in Eq.~(\ref{eq:sfermion-4D}) are 
defined by 
\begin{equation}
  \frac{1}{\tilde{m}_1^2} \equiv \frac{2}{5}\frac{1}{|m_{\cal D}|^2} 
    + \frac{3}{5}\frac{1}{|m_{\cal L}|^2},
\qquad
  \frac{1}{\tilde{m}_2^2} \equiv \frac{1}{|m_{\cal L}|^2},
\qquad
  \frac{1}{\tilde{m}_3^2} \equiv \frac{1}{|m_{\cal D}|^2},
\end{equation}
and
\begin{equation}
  \tilde{m}_1'^2 \equiv 
    \frac{2}{5}\left| \frac{k_{\cal D} F_Z}{m'_{\cal D}} \right|^2
  + \frac{3}{5}\left| \frac{k_{\cal L} F_Z}{m'_{\cal L}} \right|^2,
\qquad
  \tilde{m}_2'^2 \equiv 
    \left| \frac{k_{\cal L} F_Z}{m'_{\cal L}} \right|^2,
\qquad
  \tilde{m}_3'^2 \equiv 
    \left| \frac{k_{\cal D} F_Z}{m'_{\cal D}} \right|^2.
\end{equation}
We find that for $m_{\cal D} > m_{\cal L}$ and $m'_{\cal D}/k_{\cal D} 
> m'_{\cal L}/k_{\cal L}$, the masses of the colored particles are 
suppressed relative to those obtained from the masses of non-colored 
particles using the unified mass relations. 

The Higgs sector of our model contains the two Higgs doublets of 
the MSSM, $H_u$ and $H_d$, and a singlet $S$.  We assume the presence 
of a superpotential term of the form $W = \lambda S H_u H_d$, and 
possibly of an additional term $\delta W = (\kappa/3) S^3$.  The other 
terms in the Higgs-sector superpotential are effectively generated 
through couplings to fields in the mediator and the DSB sectors. 
Suppose there are tree-level superpotential interactions
\begin{equation}
  W = h \tilde{Q} {\cal L} H_u + \bar{h} \tilde{Q} \bar{\cal L} H_d.
\label{eq:4D-mu-gen}
\end{equation}
Then, the effective $\mu$-term is generated after integrating out the 
mediator fields as $W \simeq -h\bar{h} (\langle (\tilde{Q}\tilde{Q}) 
\rangle/m_{\cal L}) H_u H_d \equiv \mu H_u H_d$.  Since $\mu \simeq 
h \bar{h} \Lambda/4\pi$, we obtain the weak-scale $\mu$-term for 
$h \bar{h} = O(0.01\!\sim\!0.1)$.  With these values of $h$ and 
$\bar{h}$ (and taking $h \sim \bar{h}$), radiatively generated 
soft supersymmetry-breaking Higgs masses, $m_{H_u,H_d}^2 \simeq 
(h^2/16\pi^2)(|\lambda|^2 \Lambda^4 /m_{\cal L}^2) \simeq 
(h \Lambda/16\pi^2)^2$ are sufficiently small.  The $\mu B$ term 
is zero at tree level, but it is generated at radiative level. 

We now introduce a mediator field ${\cal N}({\bf 1}, {\bf 1})_0$, which 
is in the fundamental representation of $S\!P(2)$, and the superpotential
\begin{equation}
  W = \frac{m_{\cal N}}{2} {\cal N}^2 + k \tilde{Q} {\cal N} S,
\end{equation}
in parallel with Eqs.~(\ref{eq:mediators},~\ref{eq:4D-mu-gen}), where 
$m_{\cal N}$ is a mass parameter of order $4\pi \Lambda$.  Integrating out 
the ${\cal N}$ field, we obtain an effective supersymmetric mass for $S$: 
$W \simeq -(k^2/2) (\langle (\tilde{Q}\tilde{Q}) \rangle/m_{\cal N}) S^2 
\equiv (\mu_S/2) S^2$.  By choosing the values of $k$ and $m_{\cal N}$ 
appropriately, $\mu_S$ can be made to be of order the weak scale.  The 
superpotential of our Higgs sector is thus given by $W = \lambda S H_u H_d 
+ \mu H_u H_d + (\mu_S/2) S^2 + (\kappa/3) S^3$.  By shifting the $S$ 
field such that the coefficient of the $H_u H_d$ term becomes zero, we 
can write the superpotential in the form of
\begin{equation}
  W = \lambda S H_u H_d + L_S^2 S 
    + \frac{M_S}{2}S^2 + \frac{\kappa}{3}S^3,
\end{equation}
where $L_S$ and $M_S$ are mass parameters of order the weak scale. 
This is the general form of the Higgs-sector superpotential we used 
in our analysis of electroweak symmetry breaking. 

The soft supersymmetry-breaking operator of the form ${\cal L} = |S|^2$ 
is also generated through loops of ${\cal N}$ and $\tilde{Q}$, with the 
coefficient of order $(k^2/16\pi^2) (|\lambda F_Z|^2/m_{\cal N}^2) \simeq 
(k \Lambda/16\pi^2)^2$ (before the shift of $S$).  The sign of this 
coefficient is incalculable due to strong $S\!P(2)$ gauge interactions. 
After shifting $S$, we obtain soft supersymmetry-breaking interactions 
of the form 
\begin{equation}
  {\cal L}_{\rm soft} 
    = - m_{H_u}^2 |H_u|^2 - m_{H_d}^2 |H_d|^2 - m_S^2 |S|^2 
      - \Bigl( L_S^2 C_S S + {\rm h.c.} \Bigr).
\end{equation}
We thus almost reproduce the general supersymmetry-breaking terms 
in the Higgs sector used in our previous analysis.%
\footnote{If we could somehow induce the supersymmetry-breaking operator 
of the form ${\cal L} = F_S^\dagger S + {\rm h.c.}$ with the weak-scale 
coefficient, we would completely reproduce the structure of the Higgs 
sector discussed in the previous sections.}

We have seen that all the essential ingredients needed to reduce 
fine-tuning can be reproduced in our 4D theory described here.  In this 
theory, some of the parameters, especially those in the Higgs sector, 
are incalculable because of strong gauge interactions in the DSB sector. 
While we do not perform a complete analysis of electroweak symmetry 
breaking here, we expect that there is a parameter region in which 
fine-tuning is reduced in this theory or a modified/extended version.

\section{Conclusions}
\label{sec:concl}

In this paper we have constructed supersymmetric theories that do not 
suffer from significant fine-tuning in obtaining realistic electroweak 
symmetry breaking.  In these theories, supersymmetry is dynamically 
broken at relatively low scale of order $(10\!\sim\!100)~{\rm TeV}$, 
which is then transmitted to the SSM sector through standard-model 
gauge interactions.  The spectrum of superparticles does not respect 
unified mass relations because of the breaking or absence of unified 
symmetry in the supersymmetry breaking sector.  The Higgs sector of our 
theories contains a singlet field $S$ in addition to the MSSM two Higgs 
doublets $H_u$ and $H_d$, with general superpotential interactions 
given in Eq.~(\ref{eq:intro-Higgs}).  Such a superpotential can naturally 
arise through singlet fields that interact both with the $S$ field and 
the supersymmetry breaking sector.  The lightest of these singlets may 
be stable due to an unbroken $Z_2$ symmetry, and thus may constitute 
the dark matter of the universe. 

We have constructed an explicit model in warped space, and studied 
its properties.  We have analyzed electroweak symmetry breaking of 
the model in detail, performing a renormalization group analysis and 
a minimization of the Higgs potential.  This allowed us explicitly 
to demonstrate that the model allows parameter regions in which the 
fine-tuning associated with electroweak symmetry breaking, defined as 
a fractional change of the weak scale in response to fractional changes 
of fundamental parameters of the theory, is reduced to the $20\%$ level. 
This is a significant improvement over conventional supersymmetry breaking 
scenarios, which typically require fine-tuning of order $(2\!\sim\!3)\%$ 
or even worse.  The parameter region with reduced fine-tuning requires 
the superparticles to be relatively light, so that these particles 
must be seen at the LHC.  We have explicitly calculated the spectra of 
superparticles in a few sample points in parameter space, and discussed 
some of their phenomenological aspects, including neutralino phenomenology 
and pedestrian dark matter. 

We have also presented a theory constructed purely in 4D, which reproduces 
structures for the superparticle spectrum and the Higgs sector similar to 
those of our warped-space model.  This theory possesses all the essential 
features necessary to reduce fine-tuning, and thus can potentially give 
natural electroweak symmetry breaking. 

While we have worked in the context of explicit models, some of our 
analysis and considerations are general.  We hope that these results 
are useful for advancing our understanding of electroweak symmetry 
breaking in supersymmetric theories.

\section*{Acknowledgments}

Y.N. thanks Ian Hinchliffe and  Zoltan Ligeti for useful discussions. 
The work of Z.C. was supported in part by the National Science Foundation 
under grant PHY-0408954.  The work of Y.N. was supported in part by the 
Director, Office of Science, Office of High Energy and Nuclear Physics, 
of the US Department of Energy under Contract DE-AC03-76SF00098 and 
DE-FG03-91ER-40676, by the National Science Foundation under grant 
PHY-0403380, and by a DOE Outstanding Junior Investigator award. 
The work of D.T.-S. was supported by a Research Corporation Cottrell 
College Science Award.


\newpage

\begin{thebibliography}{0}

\bibitem{Dimopoulos:1981zb}
S.~Dimopoulos and H.~Georgi,
Nucl.\ Phys.\ B {\bf 193}, 150 (1981);
N.~Sakai,
Z.\ Phys.\ C {\bf 11}, 153 (1981);
S.~Dimopoulos, S.~Raby and F.~Wilczek,
Phys.\ Rev.\ D {\bf 24}, 1681 (1981).

\bibitem{Nilles:1982dy}
H.~P.~Nilles, M.~Srednicki and D.~Wyler,
Phys.\ Lett.\ B {\bf 120}, 346 (1983);
J.~M.~Frere, D.~R.~T.~Jones and S.~Raby,
Nucl.\ Phys.\ B {\bf 222}, 11 (1983);
J.~P.~Derendinger and C.~A.~Savoy,
Nucl.\ Phys.\ B {\bf 237}, 307 (1984);
J.~R.~Ellis, J.~F.~Gunion, H.~E.~Haber, L.~Roszkowski and F.~Zwirner,
Phys.\ Rev.\ D {\bf 39}, 844 (1989).

\bibitem{Nomura:2004is}
\label{Nomura:2004is:X}
Y.~Nomura, D.~Tucker-Smith and B.~Tweedie,
Phys.\ Rev.\ D {\bf 71}, 075004 (2005)
[arXiv:hep-ph/0403170].

\bibitem{Nomura:2004zs}
Y.~Nomura,
arXiv:hep-ph/0410348.

\bibitem{Barbieri:1987fn}
R.~Barbieri and G.~F.~Giudice,
Nucl.\ Phys.\ B {\bf 306}, 63 (1988).

\bibitem{Bastero-Gil:2000bw}
\label{Bastero-Gil:2000bw:X}
M.~Bastero-Gil, C.~Hugonie, S.~F.~King, D.~P.~Roy and S.~Vempati,
Phys.\ Lett.\ B {\bf 489}, 359 (2000)
[arXiv:hep-ph/0006198];
R.~Dermisek and J.~F.~Gunion,
arXiv:hep-ph/0502105.

\bibitem{Batra:2003nj}
P.~Batra, A.~Delgado, D.~E.~Kaplan and T.~M.~P.~Tait,
JHEP {\bf 0402}, 043 (2004)
[arXiv:hep-ph/0309149];
JHEP {\bf 0406}, 032 (2004)
[arXiv:hep-ph/0404251];
A.~Maloney, A.~Pierce and J.~G.~Wacker,
arXiv:hep-ph/0409127.

\bibitem{Harnik:2003rs}
R.~Harnik, G.~D.~Kribs, D.~T.~Larson and H.~Murayama,
Phys.\ Rev.\ D {\bf 70}, 015002 (2004)
[arXiv:hep-ph/0311349];
S.~Chang, C.~Kilic and R.~Mahbubani,
Phys.\ Rev.\ D {\bf 71}, 015003 (2005)
[arXiv:hep-ph/0405267].

\bibitem{Casas:2003jx}
J.~A.~Casas, J.~R.~Espinosa and I.~Hidalgo,
JHEP {\bf 0401}, 008 (2004)
[arXiv:hep-ph/0310137];
A.~Brignole, J.~A.~Casas, J.~R.~Espinosa and I.~Navarro,
Nucl.\ Phys.\ B {\bf 666}, 105 (2003)
[arXiv:hep-ph/0301121].

\bibitem{Kobayashi:2004pu}
T.~Kobayashi and H.~Terao,
JHEP {\bf 0407}, 026 (2004)
[arXiv:hep-ph/0403298];
T.~Kobayashi, H.~Nakano and H.~Terao,
arXiv:hep-ph/0502006.

\bibitem{Birkedal:2004xi}
A.~Birkedal, Z.~Chacko and M.~K.~Gaillard,
JHEP {\bf 0410}, 036 (2004)
[arXiv:hep-ph/0404197];
P.~H.~Chankowski, A.~Falkowski, S.~Pokorski and J.~Wagner,
Phys.\ Lett.\ B {\bf 598}, 252 (2004)
[arXiv:hep-ph/0407242].

\bibitem{Chacko:2004mi}
Z.~Chacko, P.~J.~Fox and H.~Murayama,
Nucl.\ Phys.\ B {\bf 706}, 53 (2005)
[arXiv:hep-ph/0406142];
see also
Y.~Nomura and D.~R.~Smith,
Phys.\ Rev.\ D {\bf 68}, 075003 (2003)
[arXiv:hep-ph/0305214].

\bibitem{Birkedal:2004zx}
A.~Birkedal, Z.~Chacko and Y.~Nomura,
Phys.\ Rev.\ D {\bf 71}, 015006 (2005)
[arXiv:hep-ph/0408329].

\bibitem{Babu:2004xg}
\label{Babu:2004xg:X}
K.~S.~Babu, I.~Gogoladze and C.~Kolda,
arXiv:hep-ph/0410085.

\bibitem{Chankowski:1998xv}
P.~H.~Chankowski, J.~R.~Ellis, M.~Olechowski and S.~Pokorski,
Nucl.\ Phys.\ B {\bf 544}, 39 (1999)
[arXiv:hep-ph/9808275];
M.~Bastero-Gil, G.~L.~Kane and S.~F.~King,
Phys.\ Lett.\ B {\bf 474}, 103 (2000)
[arXiv:hep-ph/9910506];
J.~A.~Casas, J.~R.~Espinosa and I.~Hidalgo,
JHEP {\bf 0401}, 008 (2004)
[arXiv:hep-ph/0310137].

\bibitem{Chamseddine:1982jx}
A.~H.~Chamseddine, R.~Arnowitt and P.~Nath,
Phys.\ Rev.\ Lett.\  {\bf 49}, 970 (1982);
R.~Barbieri, S.~Ferrara and C.~A.~Savoy,
Phys.\ Lett.\ B {\bf 119}, 343 (1982);
L.~J.~Hall, J.~Lykken and S.~Weinberg,
Phys.\ Rev.\ D {\bf 27}, 2359 (1983).

\bibitem{unknown:2001xx}
LEP Higgs Working Group Collaboration,
arXiv:hep-ex/0107030.

\bibitem{Okada:1990vk}
Y.~Okada, M.~Yamaguchi and T.~Yanagida,
Prog.\ Theor.\ Phys.\  {\bf 85}, 1 (1991);
J.~R.~Ellis, G.~Ridolfi and F.~Zwirner,
Phys.\ Lett.\ B {\bf 257}, 83 (1991);
H.~E.~Haber and R.~Hempfling,
Phys.\ Rev.\ Lett.\  {\bf 66}, 1815 (1991).

\bibitem{Carena:1995wu}
See, for example, 
M.~Carena, M.~Quiros and C.~E.~M.~Wagner,
Nucl.\ Phys.\ B {\bf 461}, 407 (1996)
[arXiv:hep-ph/9508343];
H.~E.~Haber, R.~Hempfling and A.~H.~Hoang,
Z.\ Phys.\ C {\bf 75}, 539 (1997)
[arXiv:hep-ph/9609331].

\bibitem{Dine:1981gu}
M.~Dine and W.~Fischler,
Phys.\ Lett.\ B {\bf 110}, 227 (1982);
Nucl.\ Phys.\ B {\bf 204}, 346 (1982);
L.~Alvarez-Gaume, M.~Claudson and M.~B.~Wise,
Nucl.\ Phys.\ B {\bf 207}, 96 (1982);
S.~Dimopoulos and S.~Raby,
Nucl.\ Phys.\ B {\bf 219}, 479 (1983).

\bibitem{Dine:1994vc}
M.~Dine, A.~E.~Nelson and Y.~Shirman,
Phys.\ Rev.\ D {\bf 51}, 1362 (1995)
[arXiv:hep-ph/9408384];
M.~Dine, A.~E.~Nelson, Y.~Nir and Y.~Shirman,
Phys.\ Rev.\ D {\bf 53}, 2658 (1996)
[arXiv:hep-ph/9507378].

\bibitem{Gherghetta:2000qt}
\label{Gherghetta:2000qt:X}
T.~Gherghetta and A.~Pomarol,
Nucl.\ Phys.\ B {\bf 586}, 141 (2000)
[arXiv:hep-ph/0003129].

\bibitem{Goldberger:2002pc}
W.~D.~Goldberger, Y.~Nomura and D.~R.~Smith,
Phys.\ Rev.\ D {\bf 67}, 075021 (2003)
[arXiv:hep-ph/0209158].

\bibitem{Chacko:2003tf}
Z.~Chacko and E.~Ponton,
JHEP {\bf 0311}, 024 (2003)
[arXiv:hep-ph/0301171].

\bibitem{Nomura:2003qb}
Y.~Nomura and D.~R.~Smith,
Phys.\ Rev.\ D {\bf 68}, 075003 (2003)
[arXiv:hep-ph/0305214].

\bibitem{Nomura:2004it}
\label{Nomura:2004it:X}
Y.~Nomura and D.~Tucker-Smith,
Nucl.\ Phys.\ B {\bf 698}, 92 (2004)
[arXiv:hep-ph/0403171].

\bibitem{Gherghetta:2000kr}
\label{Gherghetta:2000kr:X}
T.~Gherghetta and A.~Pomarol,
Nucl.\ Phys.\ B {\bf 602}, 3 (2001)
[arXiv:hep-ph/0012378];
L.~J.~Hall, Y.~Nomura, T.~Okui and S.~J.~Oliver,
Nucl.\ Phys.\ B {\bf 677}, 87 (2004)
[arXiv:hep-th/0302192];
K.~w.~Choi, D.~Y.~Kim, I.~W.~Kim and T.~Kobayashi,
arXiv:hep-ph/0301131;
Eur.\ Phys.\ J.\ C {\bf 35}, 267 (2004)
[arXiv:hep-ph/0305024];
Z.~Chacko, P.~J.~Fox and H.~Murayama,
Nucl.\ Phys.\ B {\bf 706}, 53 (2005)
[arXiv:hep-ph/0406142].

\bibitem{Anderson:1994dz}
G.~W.~Anderson and D.~J.~Castano,
Phys.\ Lett.\ B {\bf 347}, 300 (1995)
[arXiv:hep-ph/9409419].

\bibitem{Dine:1993yw}
M.~Dine and A.~E.~Nelson,
Phys.\ Rev.\ D {\bf 48}, 1277 (1993)
[arXiv:hep-ph/9303230];
K.~Agashe and M.~Graesser,
Nucl.\ Phys.\ B {\bf 507}, 3 (1997)
[arXiv:hep-ph/9704206];
A.~de Gouvea, A.~Friedland and H.~Murayama,
Phys.\ Rev.\ D {\bf 57}, 5676 (1998)
[arXiv:hep-ph/9711264].

\bibitem{Panagiotakopoulos:2000wp}
C.~Panagiotakopoulos and A.~Pilaftsis,
Phys.\ Rev.\ D {\bf 63}, 055003 (2001)
[arXiv:hep-ph/0008268];
A.~Menon, D.~E.~Morrissey and C.~E.~M.~Wagner,
Phys.\ Rev.\ D {\bf 70}, 035005 (2004)
[arXiv:hep-ph/0404184];
R.~Kitano, G.~D.~Kribs and H.~Murayama,
Phys.\ Rev.\ D {\bf 70}, 035001 (2004)
[arXiv:hep-ph/0402215].

\bibitem{Maldacena:1997re}
J.~M.~Maldacena,
Adv.\ Theor.\ Math.\ Phys.\  {\bf 2}, 231 (1998)
[Int.\ J.\ Theor.\ Phys.\  {\bf 38}, 1113 (1999)]
[arXiv:hep-th/9711200];
S.~S.~Gubser, I.~R.~Klebanov and A.~M.~Polyakov,
Phys.\ Lett.\ B {\bf 428}, 105 (1998)
[arXiv:hep-th/9802109];
E.~Witten,
Adv.\ Theor.\ Math.\ Phys.\  {\bf 2}, 253 (1998)
[arXiv:hep-th/9802150].

\bibitem{Arkani-Hamed:2000ds}
N.~Arkani-Hamed, M.~Porrati and L.~Randall,
JHEP {\bf 0108}, 017 (2001)
[arXiv:hep-th/0012148];
R.~Rattazzi and A.~Zaffaroni,
JHEP {\bf 0104}, 021 (2001)
[arXiv:hep-th/0012248].

\bibitem{Randall:1999ee}
L.~Randall and R.~Sundrum,
Phys.\ Rev.\ Lett.\  {\bf 83}, 3370 (1999)
[arXiv:hep-ph/9905221].

\bibitem{Pomarol:2000hp}
A.~Pomarol,
Phys.\ Rev.\ Lett.\  {\bf 85}, 4004 (2000)
[arXiv:hep-ph/0005293].

\bibitem{Nomura:2005qg}
Y.~Nomura and B.~Tweedie,
arXiv:hep-ph/0504246.

\bibitem{Marti:2001iw}
D.~Marti and A.~Pomarol,
Phys.\ Rev.\ D {\bf 64}, 105025 (2001)
[arXiv:hep-th/0106256].

\bibitem{'tHooft:1973jz}
G.~'t Hooft,
Nucl.\ Phys.\ B {\bf 72}, 461 (1974);
E.~Witten,
Nucl.\ Phys.\ B {\bf 160}, 57 (1979).

\bibitem{Burdman:2003ya}
G.~Burdman and Y.~Nomura,
Phys.\ Rev.\ D {\bf 69}, 115013 (2004)
[arXiv:hep-ph/0312247].

\bibitem{Manohar:1983md}
A.~Manohar and H.~Georgi,
Nucl.\ Phys.\ B {\bf 234}, 189 (1984);
H.~Georgi and L.~Randall,
Nucl.\ Phys.\ B {\bf 276}, 241 (1986);
Z.~Chacko, M.~A.~Luty and E.~Ponton,
JHEP {\bf 0007}, 036 (2000)
[arXiv:hep-ph/9909248];
see also
Y.~Nomura,
Phys.\ Rev.\ D {\bf 65}, 085036 (2002)
[arXiv:hep-ph/0108170].

\bibitem{Kane:1992kq}
G.~L.~Kane, C.~F.~Kolda and J.~D.~Wells,
Phys.\ Rev.\ Lett.\  {\bf 70}, 2686 (1993)
[arXiv:hep-ph/9210242];
M.~Masip, R.~Munoz-Tapia and A.~Pomarol,
Phys.\ Rev.\ D {\bf 57}, 5340 (1998)
[arXiv:hep-ph/9801437].

\bibitem{Lopez:1991aw}
J.~L.~Lopez and D.~V.~Nanopoulos,
Phys.\ Lett.\ B {\bf 266}, 397 (1991);
P.~H.~Chankowski, J.~R.~Ellis, M.~Olechowski and S.~Pokorski,
Nucl.\ Phys.\ B {\bf 544}, 39 (1999)
[arXiv:hep-ph/9808275].

\bibitem{Heinemeyer:1998yj}
S.~Heinemeyer, W.~Hollik and G.~Weiglein,
Comput.\ Phys.\ Commun.\  {\bf 124}, 76 (2000)
[arXiv:hep-ph/9812320].

\bibitem{Eidelman:2004wy}
S.~Eidelman {\it et al.}  [Particle Data Group Collaboration],
Phys.\ Lett.\ B {\bf 592}, 1 (2004).

\bibitem{Borzumati:1998tg}
F.~M.~Borzumati and C.~Greub,
Phys.\ Rev.\ D {\bf 58}, 074004 (1998)
[arXiv:hep-ph/9802391];
Phys.\ Rev.\ D {\bf 59}, 057501 (1999)
[arXiv:hep-ph/9809438];
M.~Ciuchini, G.~Degrassi, P.~Gambino and G.~F.~Giudice,
Nucl.\ Phys.\ B {\bf 527}, 21 (1998)
[arXiv:hep-ph/9710335].

\bibitem{Izawa:1997gs}
K.~I.~Izawa, Y.~Nomura, K.~Tobe and T.~Yanagida,
Phys.\ Rev.\ D {\bf 56}, 2886 (1997)
[arXiv:hep-ph/9705228];
K.~I.~Izawa, Y.~Nomura and T.~Yanagida,
Phys.\ Lett.\ B {\bf 452}, 274 (1999)
[arXiv:hep-ph/9901345].

\bibitem{Izawa:2005yf}
K.~I.~Izawa and T.~Yanagida,
arXiv:hep-ph/0501254.

\bibitem{Chacko:1998si}
Z.~Chacko, M.~A.~Luty and E.~Ponton,
JHEP {\bf 9812}, 016 (1998)
[arXiv:hep-th/9810253].

\bibitem{Agashe:1997kn}
K.~Agashe and M.~Graesser,
Nucl.\ Phys.\ B {\bf 507}, 3 (1997)
[arXiv:hep-ph/9704206].

\end{thebibliography}
\end{document}